\documentclass[journal=jctcce,manuscript=article,layout=twocolumn]{achemso}
\usepackage{algorithm}
\usepackage{algpseudocode}
\usepackage[version=4]{mhchem} 
\usepackage{amsfonts}
\usepackage{amssymb,amsmath}
\usepackage{enumerate}
\usepackage{bbm}
\usepackage{color}
\definecolor{revisiongreen}{rgb}{0,0.5,0}
\usepackage{graphicx}
\usepackage{placeins}
\usepackage{bm} 
\usepackage{mathtools}
\usepackage{ulem}
\usepackage{subcaption}
\usepackage{upgreek}
\usepackage{siunitx} 
\usepackage{multirow}
\usepackage{booktabs}
\usepackage[export]{adjustbox}
\usepackage{makecell}
\usepackage{hyperref} 

\usepackage[noabbrev]{cleveref} 
\crefformat{equation}{Eq.~(#2#1#3)}
\crefformat{figure}{Fig.~#2#1#3}
\crefformat{table}{Table #2#1#3}

\usepackage{etoolbox} 
\pretocmd{\cite}{~}{}{}
\pretocmd{\ref}{~}{}{}

\usepackage{xparse, xspace}
\NewDocumentCommand\bdgrnn{g}{%
  \textrm{BDG}%
  \IfNoValueTF{#1}{}{$(#1)$}%
  \textrm{--RNN}%
  \xspace%
}

\let\Psi\relax
\let\Phi\relax
\let\Psi\varPsi
\let\Phi\varPhi

\newcommand{\rNc}{N_\mathrm{C}}
\newcommand{\rNo}{N_\mathrm{O}}
\newcommand{\rNv}{N_\mathrm{V}}
\newcommand{\rX}{\mathrm{X}}

\newcommand{\dST}{\Delta E_{\mathrm{ST}}}
\newcommand{\dSpinSquare}{\Delta \langle\hat{S}^2\rangle}

\title{
Spin-Adapted Spin-Flip Tamm-Dancoff Approximation (XSF-TDA) for Singlet-Triplet Gaps in MR-TADF and INVEST Emitters
}

\author{Hewang Zhao}
\affiliation{Key Laboratory of Theoretical and Computational Photochemistry, Ministry of Education, College of Chemistry, Beijing Normal University, Beijing, 100875, China}
\alsoaffiliation{Institute for Advanced Study, Beijing Normal University, Beijing, 100875, China}
\author{Hongbo Wang}
\affiliation{Key Laboratory of Theoretical and Computational Photochemistry, Ministry of Education, College of Chemistry, Beijing Normal University, Beijing, 100875, China}
\alsoaffiliation{Institute for Advanced Study, Beijing Normal University, Beijing, 100875, China}
\author{Liuxin Huang}
\affiliation{College of Chemistry and Molecular Engineering, Peking University, Beijing 100871, China}
\author{Yunlong Xiao}\email{xiaoyl@pku.edu.cn}
\affiliation{College of Chemistry and Molecular Engineering, Peking University, Beijing 100871, China}
\author{Zhendong Li}\email{zhendongli@bnu.edu.cn}
\affiliation{Key Laboratory of Theoretical and Computational Photochemistry, Ministry of Education, College of Chemistry, Beijing Normal University, Beijing, 100875, China}
\alsoaffiliation{Institute for Advanced Study, Beijing Normal University, Beijing, 100875, China}

\begin{document}

%
%
%
%
%

\begin{tocentry}
%
%
%
%
%
\includegraphics[]{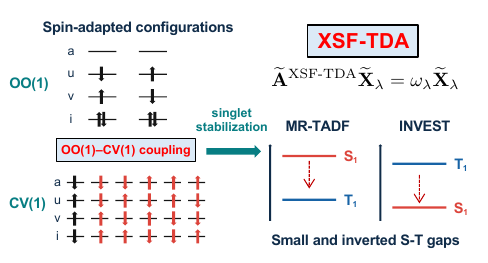}
\end{tocentry}

\begin{abstract}
Reliable singlet--triplet gaps are central to the design of multiple-resonance thermally activated delayed-fluorescence (MR-TADF) and inverted singlet--triplet (INVEST) emitters, yet conventional time-dependent density functional theory (TD-DFT) and its Tamm--Dancoff approximation (TDA) often fail to capture the small correlation-driven energy shifts that determine the magnitude and sign of $\dST$. In this work, we assess the recently introduced spin-adapted TDA for spin-flip-down excitations (XSF-TDA) [Mol. Phys. e2631735 (2026)] as a practical approach to this problem. Starting from a high-spin restricted open-shell Kohn--Sham reference, XSF-TDA generates spin-adapted low-spin roots arising from open--open (OO), core--open (CO), open--virtual (OV),
and core--virtual (CV) types of spin-flip-down excitations. Across four application sets comprising 243 benchmark cases, XSF-TDA substantially reduces the systematic positive bias observed for TDA and unrestricted spin-flip TDA (USF-TDA). With the B3LYP functional, XSF-TDA reduces the equal-set-weighted macro-MAE relative to USF-TDA from 0.252 eV (0.158 eV) to 0.102~eV (0.056~eV) with the ALDA0 (multicollinear) kernel. Its accuracy is competitive with state-specific $\Delta$DFT and quasiparticle-energy DFT (QE-DFT), while being able to provide deeper mechanistic insight through the response vector.
In particular, spin-adapted CV configurations provide selected higher-excitation character that strengthens the OO--CV dynamical-spin-polarization channel, which preferentially stabilizes the singlet states. 
We also identify an INVEST emitter with appreciable OO--OV mixing, suggesting a possible additional contribution to singlet stabilization when the CV contribution is small. The XSF-TDA method therefore provides practical singlet--triplet gap predictions together with interpretable state analysis for the screening and mechanistic understanding of MR-TADF and INVEST emitters. 
\end{abstract}

\section{Introduction}\label{introduction}
Thermally activated delayed fluorescence (TADF) enables organic light-emitting diodes to harvest both singlet and triplet excitons by reverse intersystem crossing (rISC), provided that the lowest singlet--triplet gap, $\dST=E(\mathrm{S}_1)-E(\mathrm{T}_1)$, is sufficiently small.\cite{Hiroki2012Highly} Multiple-resonance TADF\cite{hatakeyama2016ultrapure} (MR-TADF) introduced a complementary molecular-design principle: alternating electron-rich and electron-deficient atoms in a rigid $\pi$ framework localizes the frontier orbitals on adjacent atomic sites, simultaneously reducing their exchange interaction and suppressing the structural relaxation that broadens conventional donor--acceptor emission. The seminal report by Hatakeyama and co-workers\cite{hatakeyama2016ultrapure} demonstrated this multiple-resonance effect in ultrapure-blue TADF molecules, establishing the foundation of the modern MR-TADF field. The resulting combination of narrow emission, high color purity, and efficient triplet harvesting has stimulated extensive experimental and computational development.\cite{hall2022,kunze2025}

An even more demanding regime is encountered in inverted singlet--triplet (INVEST) molecules, for which S$_1$ lies below T$_1$ ($\dST<0$), thereby eliminating, in principle, the endothermic energy-gap penalty for reverse intersystem crossing (rISC).
The predicted inversions of heptazine, cyclazine, and related compounds\cite{ehrmaier2019,de2019,2023heptazine} and their subsequent experimental realization\cite{0Delayed} have made reliable prediction of both small positive and negative gaps an important problem in emitter screening.\cite{deSousa2021,deSousa2022,Bedongni2024,D5TC01873K,D5TC01155H} Although MR-TADF and INVEST are not synonymous material classes, they share a difficult electronic-structure regime: reduced frontier-orbital exchange first creates a near-degeneracy, after which singlet-selective dynamical spin polarization (DSP), orbital relaxation, and dynamic correlation determine the magnitude and sometimes the sign of $\dST$.\cite{drwal2023spin,peng2025,omar2023,hall2022}

This correlation balance explains why the problem is not solved merely by choosing an appropriate
exchange-correlation (XC) functional in conventional time-dependent density functional
theory (TD-DFT)\cite{PhysRevLett.52.997,PhysRevLett.76.1212} and its Tamm-Dancoff approximation (TDA).\cite{HIRATA1999291}
Important DSP pathways contain double- and higher-excitation character relative to a closed-shell determinant, whereas conventional adiabatic linear-response TD-DFT and TDA lack an explicit description of double excitations.\cite{maitra2004double} This limitation contributes to the positive singlet--triplet gap bias observed for many INVEST systems.\cite{2023heptazine,drwal2023spin} Wave-function methods such as coupled cluster, algebraic-diagrammatic construction, and multireference perturbation theory can recover more of this physics, but their cost limits systematic screening of realistic emitters.\cite{C1966On,CHRISTIANSEN1995409,Schirmer1982Beyond,trofimov1995adc,angeli2001nevpt2,2004Calibration} Double-hybrid response methods introduce perturbative correlation,\cite{double2022,alipour2022,D5TC01799H,yang2026} while state-specific $\Delta$DFT and quasiparticle-energy DFT (QE-DFT) can benefit from separately optimized states.\cite{gilbert2008,kowalczyk2013,hait2021,mei2019,haiduke2018,kunze2024,kunze2025,li2026} The latter approaches, however, do not by themselves provide one response-state manifold for transition moments and related excited-state properties. A useful complementary target is therefore a response method that remains near TD-DFT cost but admits the configurations controlling correlation-driven gaps.

Spin-flip TD-DFT approaches this target by generating low-spin states from a high-spin reference.\cite{shao2003spin,krylov2001size,casanova2020spin,herbert2022spin} This compact construction already captures important multiconfigurational character, but ordinary spin-unrestricted SF-TDA (USF-TDA) can suffer from severe, state-dependent spin contamination.\cite{li2012theoretical,li2010Spin} Several partially or completely spin-adapted variants have therefore been proposed.\cite{zhang2015spin,chibueze2025spin,lee2018eliminating,lee2019efficient,2025spinadapted} A thorough analysis of different spin-adapted spin-flip-down
methods is presented in Ref. \cite{2025spinadapted}. Here, we only summarize the major distinctions. Specifically, spin-flip-down single excitations from a high-spin reference can be classified into four categories: OO(1), CO(1), OV(1), and CV(1),
where C, O, and V refer to closed-shell, open-shell, and vacant-shell orbitals, respectively, and (1) indicates the excitation is an triplet excitation.
The mixed-reference SF-TDA\cite{lee2018eliminating} (MRSF-TDA) spin-adapts the 
OO(1), CO(1), and OV(1) sectors only for a triplet reference.
Both the spin-adapted spin-flip DFT (SA-SF-DFT)\cite{zhang2015spin} and
the recently proposed XSF-TDA\cite{2025spinadapted} spin-adapt all four sectors, including the CV(1) configurations most directly connected with DSP. However, XSF-TDA constructs a Hermitian response equation following our previous construction for spin-conserving excitations in X-TD-DFT\cite{li2010Spin,li2011spin,li2011spin2}, whereas the eigenvalue equation in SA-SF-DFT is non-Hermitian.

The original XSF-TDA study established the method on spin-flip benchmarks, atomic degeneracies, and bond-breaking curves,\cite{2025spinadapted} and included a pilot application to the ten-molecule triangulene benchmark of Loos and co-workers.\cite{2023heptazine} For that benchmark, XSF-TDA/B3LYP and XSF-TDA/PBE0 with aug-cc-pVTZ achieved mean absolute errors (MAEs) of 0.07 and 0.10~eV, respectively, and markedly improved upon conventional TDA, USF-TDA, and the tested MRSF-TDA protocols.\cite{2025spinadapted}
Those results were encouraging but deliberately limited in chemical scope. The present study moves from that ten-molecule proof of concept to systematic validation on 243 application cases: the NAH159 set\cite{garner2024,kunze2024}, 35 MR-TADF benchmark\cite{hall2022,kunze2025}, four experimentally characterized heptazine derivatives\cite{0Delayed}, and the 45 INVEST molecules derived by Pang et al.\cite{peng2025}.
This broader design makes it possible to ask not only whether XSF-TDA works for prototypical inverted gaps, but also when it transfers across calculated and experimental references and chemically realistic MR-TADF structures. Besides, the previous XSF-TDA uses the ALDA0 kernel\cite{li2012theoretical,li2016critical} to overcome 
numerical instabilities encountered in the spin-flip noncollinear kernel for XC functionals
involving generalized gradient approximation (GGA). In this work, we extend it
to support the multicollinear kernel\cite{wang2025zero}, which contains GGA contribution to the spin-flip kernel, but is free from numerical instabilities.

The enlarged dataset also enables a more useful mechanistic analysis for molecular screening. In XSF-TDA calculations for $\dST$, the $M_S=+1$ component of the T$_1$ state is used as the reference, and $\dST$ between S$_1$ and T$_1$ is directly obtained as the second lowest eigenvalue. Hence, the amplitudes for configurations arising from different classes of excitations can
be used to analyze different mechanisms that leads to small or negative $\dST$. 
This provides a common response-vector framework for analyzing the configuration mixing associated with small or negative $\dST$, which is advantageous over other state-specific 
DFT or TD-DFT based methods that only correct $\dST$.
For instance, in conventional CIS(D)-corrected double-hybrid TD-DFT/TDA, the perturbative contribution is added to the excitation energy without rediagonalizing the underlying response problem, so its response amplitudes are unchanged by this energy correction.\cite{grimme2007doublehybrid} State-specific $\Delta$DFT and QE-DFT can also support electronic-structure analysis, but do not provide the same decomposition into spin-adapted OO(1), CO(1), OV(1), and CV(1) response amplitudes. The results below suggest a qualitative association between larger CV(1) weights and smaller $\dST$, particularly within NAH159, consistent with DSP-like singlet stabilization.
Besides, XSF-TDA also identifies an interesting INVEST molecule for which the CV(1) weight is small but the OV(1) weight is appreciable. This case motivates further examination of OO(1)--OV(1) mixing as a possible additional contribution to singlet stabilization that has, to the best of our knowledge, received limited attention in previous MR-TADF/INVEST studies.

Accordingly, this work aims to answer three application-oriented questions: (i) how accurately does XSF-TDA predict positive, near-zero, and inverted gaps across realistic MR-TADF and INVEST sets; (ii) which functional/kernel combinations provide a sensible starting protocol; (iii) how can CV(1), OV(1), and CO(1) amplitudes be used to recognize the electronic mechanism of small or inverted gaps.
We provide an open-source implementation of XSF-TDA 
in the \textsc{XTDDFT} package,\cite{XTDDFTcode} and
the results of this work provides a guide to apply XSF-TDA
to MR-TADF and INVEST systems.

The remaining part of this paper is organized as follows:
Section~\ref{sec:theory} presents the XSF-TDA method and
its extension to the multicollinear kernel, as 
well as an analysis of the possible singlet stabilization channels. Section~\ref{sec:comp} summarizes the computational details. Section~\ref{sec:result} first calibrates the multicollinear extension, then tests transferability across MR-TADF and INVEST sets and shows how the weights of different classes of excitations can be used to interpret the negative ST gaps. 
Conclusions are given in Section~\ref{sec:conclusion}.

\section{Theory}\label{sec:theory}
\subsection{XSF-TDA response equation and spin-adapted manifold}
\begin{figure*}[!tp]
\centering
\includegraphics[width=0.8\linewidth]{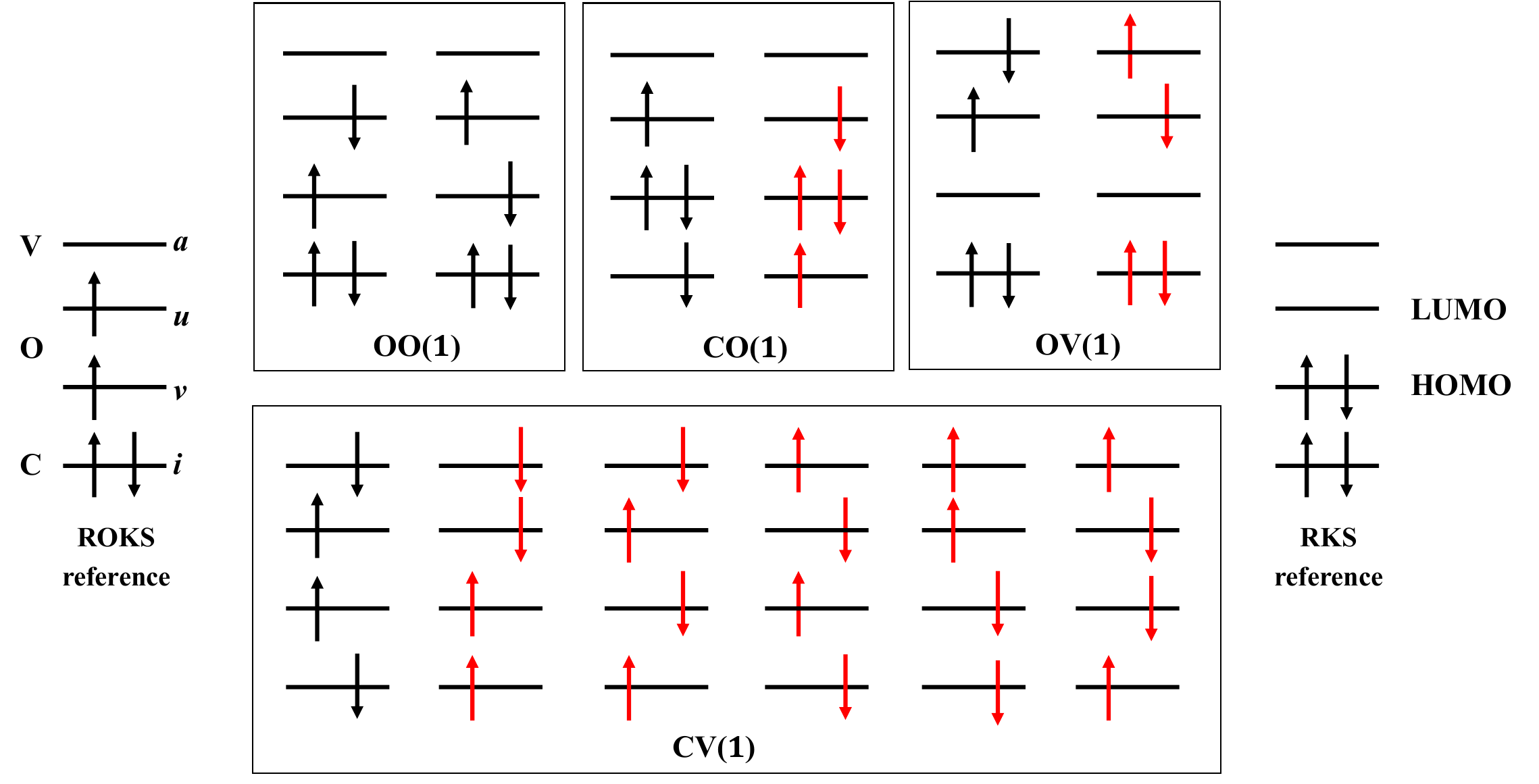}
\caption{
Illustration for a high-spin triplet ROKS reference and the four different classes of spin-flip-down excitations.
The excited configurations in black are generated from single spin-flip-down excitations
from the ROKS reference, while the other configurations highlighted in red
are higher excitations with respect to the ROKS reference, but are required by spin adaptation.
Conventional USF-TDA misses these configurations, but XSF-TDA can describe their effects via the spin-adapted correction term\cite{2025spinadapted}. For closed-shell systems considered in this work,
their ground states are all dominated by the closed-shell reference, which is used in normal TD-DFT/TDA calculations. 
The lowest unoccupied molecular orbital (LUMO) and highest occupied molecular orbital (HOMO) are defined with respect to the RKS reference.
}
\label{fig:cv_configurations}
\end{figure*}

We use the high-spin component of the lowest triplet (T$_1$) state as reference in spin-flip calculations, such that the closed-shell ground state (S$_0$) is obtained as the lowest solution of spin-flip TD-DFT/TDA calculations.
This is illustrated in Fig. \ref{fig:cv_configurations}, with the restricted open-shell Kohn-Sham (ROKS) reference shown in the left.
Throughout this work, $i,j$ label doubly occupied (core, C) orbitals, $t,u,v,w$ label singly occupied (open, O) orbitals, $a,b$ label unoccupied (virtual, V) orbitals, and $p,q,r,s$ denote general orbitals. A bar denotes a $\beta$ spin orbital. 

Starting from the ROKS determinant with spin $S_i$ (=1 in this work), the XSF-TDA eigenvalue equation for the spin-pure $S_f=S_i-1$ states with excitation energies $\omega_\lambda$ and amplitudes $\widetilde{\mathbf X}_{\lambda}$ reads\cite{2025spinadapted}
\begin{equation}
\widetilde{\mathbf A}^{\mathrm{XSF\textrm{-}TDA}}
\widetilde{\mathbf X}_{\lambda}
=\omega_{\lambda}\widetilde{\mathbf X}_{\lambda},
\label{eq:AX}
\end{equation}
where the projected response matrix $\widetilde{\mathbf A}^{\mathrm{XSF\textrm{-}TDA}}$ is given by
{\small\begin{equation}
\widetilde{\mathbf A}^{\mathrm{XSF\textrm{-}TDA}}
=\mathbf V^{T}\!\left[
\mathbf{A}^{\mathrm{USF\textrm{-}TDA}}
+g_\rX\Delta\mathbf A^{\mathrm{TCIS}}(S_i)
\right]\!\mathbf V.
\label{eq:XSFTDA}
\end{equation}}Here, $\mathbf{A}^{\mathrm{USF\textrm{-}TDA}}$ is the response matrix in conventional spin-unrestricted spin-flip TDA\cite{shao2003spin}
 (USF-TDA), viz.,
\begin{align} 
&\mathbf{A}^{\mathrm{USF\textrm{-}TDA}}\mathbf{X}_\lambda
=\omega_\lambda\mathbf{X}_\lambda,\label{eq:USF-TDA}\\
&A_{\bar p q,\bar r s}^{\mathrm{USF\textrm{-}TDA}}
=\delta_{qs}f_{pr}^{\beta}-\delta_{pr}f_{sq}^{\alpha}
+K_{\bar p q,\bar r s},
\label{eq:SFTDA_amat}
\end{align}
where $f^{\sigma}_{pq}$ is the Kohn--Sham Fock matrix for spin $\sigma$, and $K_{\bar p q,\bar r s}$ is the coupling kernel, whose explicit forms will be discussed later.

Let $\rNo=2S_i$ be the number of open-shell orbitals in the ROKS reference. USF-TDA in Eq. \eqref{eq:USF-TDA} involves $\rNo$ spin-flip-down excitations generated within the same spatial orbitals, denoted by the O$_1$O$_1$(1) type. The equal-superposition of such excitations is equivalent to the action of
the spin step-down operator $\hat{S}^-$, and hence will 
reproduce the reference state with spin $S_i$, rather than an excited state with spin $S_f=S_i-1$. 
The role of $\mathbf V$ is simply to project out the equal-superposition of such excitations, and the relevant sub-block 
$\mathbf{V}^{\mathrm{O_1O_1}}$ is an $\rNo\times (\rNo-1)$ matrix chosen to be\cite{2025spinadapted,1979Spin}
{\footnotesize\begin{equation}
    \mathbf{V}^{\mathrm{O_1O_1}}=\begin{bmatrix}
        \frac{\rNo-1}{\sqrt{\rNo(\rNo-1)}}& 0 & \dots & 0\\
			-\frac{1}{\sqrt{\rNo(\rNo-1)}}& \frac{\rNo-2}{\sqrt{(\rNo-1)(\rNo-2)}} & \dots&0\\
			\vdots&\vdots&\ddots&\vdots\\
			-\frac{1}{\sqrt{\rNo(\rNo-1)}}& -\frac{1}{\sqrt{(\rNo-1)(\rNo-2)}}&\dots&\frac{1}{\sqrt{2}}\\
			-\frac{1}{\sqrt{\rNo(\rNo-1)}}&-\frac{1}{\sqrt{(\rNo-1)(\rNo-2)}}&\dots&-\frac{1}{\sqrt{2}}
    \end{bmatrix},\label{eq:Vo1o1}
\end{equation}}
which clearly removes the equal superposition, and produces
an orthonormal basis for the $S_f=S_i-1$ excited configurations generated from the $\mathrm{O_1O_1}$ type of excitations.

A summary of the final orthonormal spin-adapted excited configurations in XSF-TDA, generated from all classes of spin-flip excitations, is given in Table \ref{tab:orthoconfs}. Except for the OO(1) type, other types of spin-adapted
configurations involve doubles and triples with respect to the ROKS reference,
which are missing in USF-TDA as illustrated in Fig. \ref{fig:cv_configurations}.
The effects of these missing configurations are reflected in the second term of Eq. \eqref{eq:XSFTDA}.
The spin-adapted correction $\Delta\mathbf A^{\mathrm{TCIS}}(S_i)$ is derived from tensor configuration interaction singles (TCIS)\cite{2025spinadapted}. 
In the case of density functional calculations, since certain correlation effects have already
been taken into account in $\mathbf{A}^{\mathrm{USF\textrm{-}TDA}}$, an empirical 
coefficient $g_\rX\in[0,1]$ is introduced to attenuate the magnitude of this correction 
in order to reduce double counting of correlation. Its specific value should depend
on the XC functional and types of spin-flip kernel in use, as discussed below.
In contrast to an a posteriori CIS(D) energy correction in conventional double-hybrid TD-DFT/TDA,\cite{grimme2007doublehybrid} $\Delta\mathbf A^{\mathrm{TCIS}}(S_i)$ modifies the response matrix and therefore changes the response vectors and the weights of the different configuration classes.

\begin{table*}[!tp]
  \centering
  \caption{\raggedright Orthonormal spin-adapted excited configurations in XSF-TDA. 
  The matrix elements $V_{t,k}$
  ($k=1,\cdots,\rNo-1$) are given in Eq.~\eqref{eq:Vo1o1}.
  $\rNc$, $\rNo$, and $\rNv$ refer to the numbers of closed-shell,
  open-shell, and vacant-shell orbitals, respectively.}
  \scalebox{0.8}{
  \begin{tabular}{ccc}
    \hline\hline
    type & dimension & expanded form for the high-spin component \\
    \hline
    CV(1) & $\rNc\rNv$ &
    $\displaystyle
    \sqrt{\frac{2S_{i}-1}{2S_{i}+1}}
    \left(
    |\Psi_{i}^{\bar a}\rangle
    -\frac{1}{2S_{i}}\sum_{t}
    \left(
    |\Psi_{it}^{a\bar t}\rangle
    -|\Psi_{\bar it}^{\bar a\bar t}\rangle
    \right)
    \right)
    -\frac{1}{2S_{i}\sqrt{(2S_{i}+1)(2S_{i}-1)}}
    \sum_{tu}|\Psi_{\bar iut}^{a\bar u\bar t}\rangle$ \\
    CO(1) & $\rNc\rNo$ &
    $\displaystyle
    |\Psi_{i}^{\bar u}\rangle
    \sqrt{\frac{2S_{i}-1}{2S_{i}}}
    +\sum_{t}|\Psi_{\bar it}^{\bar u\bar t}\rangle
    \frac{1}{\sqrt{2S_{i}(2S_{i}-1)}}$
    \\
    OV(1) & $\rNo\rNv$ &
    $\displaystyle
    |\Psi_{u}^{\bar a}\rangle
    \sqrt{\frac{2S_{i}-1}{2S_{i}}}
    -\sum_{t}|\Psi_{ut}^{a\bar t}\rangle
    \frac{1}{\sqrt{2S_{i}(2S_{i}-1)}}$
    \\
    O$_1$O$_2$(1) & $\rNo(\rNo-1)$ &
    $|\Psi^{\bar t}_{u}\rangle\quad(t\ne u)$
    \\
    O$_1$O$_1$(1) & $\rNo-1$ &
    $\displaystyle
    \sum_{t}|\Psi_{t}^{\bar t}\rangle V_{t,k}$
    \\
    \hline\hline
  \end{tabular}}
  \label{tab:orthoconfs}
\end{table*}

\FloatBarrier
\subsection{Spin-flip kernels and attenuation of the spin-adapted correction}\label{sec:kernels}
Throughout this work, we focus on GGA and global hybrid functionals.
Then, the spin-flip coupling in Eq. \eqref{eq:SFTDA_amat} 
contains the exact exchange and an XC contribution
\begin{equation}
K_{\bar p q,\bar r s}
=-c_\rX(\bar p\bar r|sq)
+[K_{\mathrm{XC}}]_{\bar p q,\bar r s},\\
\label{eq:K}
\end{equation}
where $c_\rX$ is the amount of Hartree-Fock exchange in a given  functional. 
The XC kernel depends on the specific parameterization of XC functionals. For collinear parameterization, this term simply vanishes\cite{shao2003spin}. For noncollinear parameterization, it takes the following Wang-Ziegler form\cite{wang2004time}
\begin{equation}
[K_{\mathrm{XC}}]_{\bar p q,\bar r s}
=\int\!\psi_{\bar p}(\mathbf r)\psi_q(\mathbf r)
f_{\mathrm{xc}}^{+-}(\mathbf r)
\psi_s(\mathbf r)\psi_{\bar r}(\mathbf r)
\,d\mathbf r,
\end{equation}
where
\begin{equation}
f_{\mathrm{xc}}^{+-}(\mathbf r)
=
\frac{v_{\mathrm{xc}}^{\alpha}(\mathbf r)-v_{\mathrm{xc}}^{\beta}(\mathbf r)}
{\rho_{\alpha}(\mathbf r)-\rho_{\beta}(\mathbf r)},
\end{equation}
where $v_{\mathrm{xc}}^\sigma$ denotes the semilocal exchange--correlation potential. Therefore, it enables SF-TD-DFT with pure density functionals. Unfortunately,
for GGA functionals, the kernel can diverge at the region where $\rho_{\alpha}(\mathbf r)=\rho_{\beta}(\mathbf r)$ but the density gradient is nonzero\cite{li2016critical}. Inspired by the adiabatic local density approximation (ALDA), a simple remedy is given by 
\begin{equation}
f_{\mathrm{xc}}^{+-,\mathrm{ALDA0}}(\mathbf r)
=
\left.
f_{\mathrm{xc}}^{+-}(\mathbf r)
\right|_{\nabla\rho_{\alpha}=\nabla\rho_{\beta}=\mathbf 0},
\label{eq:alda0}
\end{equation}
where the gradient is set to zero in evaluating the XC kernel, leading to well-behaved expressions. This is referred to as the ALDA0 kernel\cite{li2016critical}, and has been successfully applied
in spin-flip-up and spin-flip-down excitations\cite{li2016critical}.
In our previous work, we choose the following linear interpolation 
for a global hybrid\cite{2025spinadapted}
\begin{equation}
g_\rX^{\mathrm{ALDA0}}
=(1-g_{\mathrm{LDA}})c_\rX+g_{\mathrm{LDA}},
\quad g_{\mathrm{LDA}}=0.3,
\label{eq:rX}
\end{equation}
which recovers the TCIS limit at $c_\rX=1$, while retaining a nonzero spin-adaptation correction for general case.

The deliberate omission of gradient contributions makes ALDA0 a robust baseline rather than a complete noncollinear extension of a semilocal functional.
In this work, we also consider the multicollinear (MCOL) construction\cite{li2023noncollinear,wang2025zero}, which instead extends the underlying collinear functional by averaging over global spin-projection directions. It preserves the collinear limit and global spin-rotation invariance, has well-defined functional derivatives, and satisfies the global zero-torque condition while permitting local exchange--correlation torques. At a collinear reference, its transverse contribution for spin-flip excitations can be written
{\small
\begin{equation}
\begin{aligned}
[K_{\mathrm{XC}}^{\perp,\mathrm{MCOL}}]_{\bar p q,\bar r s}
&=2\iint\!\psi_{\bar p}(\mathbf r)\psi_q(\mathbf r)
\left.
\frac{\delta^2E_{\mathrm{xc}}^{\mathrm{MCOL}}}
{\delta m_x(\mathbf r)\delta m_x(\mathbf r')}
\right|_{\mathbf m=(0,0,m_z)}\\[-2pt]
&\quad\times\psi_s(\mathbf r')\psi_{\bar r}(\mathbf r')
\,d\mathbf r\,d\mathbf r'.
\end{aligned}
\label{eq:multicollinear}
\end{equation}}For semilocal functionals, the same transverse kernel can equivalently be constructed from the longitudinal collinear kernel through the magnetization-path average\cite{wang2025zero}
\begin{equation}
\mathbf K_{\mathrm{xc}}^{\perp,\mathrm{MCOL}}[D_z]
=2\int_0^1\mathbf K_{\mathrm{xc}}^{zz}[tD_z]\,dt,
\label{eq:mc_path_average}
\end{equation}
where $D_z(\mathbf r)=\rho_\alpha(\mathbf r)-\rho_\beta(\mathbf r)$ denotes the collinear real-space spin-magnetization density. This expression reveals why the multicollinear kernel remains nonzero for pure functionals, and can retain semilocal density-gradient information without the ALDA0 substitution.

Since the MCOL kernel include additional GGA contributions, $g_{\rX}$ needs to be further scaled down in order to reduce double counting of correlation. To this end, we introduce an empirical scaling factor
$s_\rX\in[0,1]$ for GGA and global hybrid functionals, such that
\begin{equation}
g_\rX^{\mathrm{MCOL}}
=
s_\rX
g_\rX^{\mathrm{ALDA0}}.\label{eq:rX_mc}
\end{equation}
The form of the scaling factor is calibrated in Sec. \ref{sec:kernel_results}.


\subsection{Singlet stabilization channels}\label{sec:dsp}

To discuss $\dST$ in XSF-TDA, we restrict to the special case $c_\rX=1$ (i.e., TCIS\cite{2025spinadapted}) for clarity. For comparison, in conventional CIS, the singlet-triplet gap
between the lowest singlet and triplet excited states, $\dST=E(\mathrm{S}_1)-E(\mathrm{T}_1)$,
is obtained from the difference between the lowest CIS eigenvalues for singlet and triplet, respectively. It can be proved that $\dST$ is always positive\cite{de2019}. In XSF-TDA,
$\dST$ is directly obtained as the second lowest eigenvalue from Eq. \eqref{eq:AX}, while the
lowest eigenvalue is given for the T$_1\to $ S$_0$ transition. 

For the triplet reference, we first restrict the OO(1) space to the two same-orbital spin-flip configurations, $u\to \bar{u}$ and $v\to \bar{v}$, which form the O$_1$O$_1$(1) subspace. The O$_1$O$_2$(1) configurations are excluded from this illustrative $2\times2$ block. In this subspace, Eq. \eqref{eq:SFTDA_amat} gives
\begin{eqnarray}
&&\begin{bmatrix}
f_{uu}^\beta - f_{uu}^\alpha - (uu|uu) & -(uv|vu) \\
-(vu|uv) & f_{vv}^\beta - f_{vv}^\alpha - (vv|vv)
\end{bmatrix} \nonumber\\
&=&
\begin{bmatrix}
(uv|vu) & -(uv|vu) \\
-(vu|uv) & (vu|uv) \\
\end{bmatrix},
\end{eqnarray}
where $f_{uu}^\beta - f_{uu}^\alpha=(uu|uu)+(uv|vu)$ has been used for the ROHF reference.
Transforming this matrix using $\mathbf{V}$ in Eq. \eqref{eq:Vo1o1} for $S_i=1$ gives the matrix element in XSF-TDA for the S$_1$ state as
\begin{eqnarray}
\omega_{\mathrm{S}_1}^{(0)}=
\widetilde{A}^{\mathrm{XSF\textrm{-}TDA}}_{\mathrm{S}_1,\mathrm{S}_1}
=2(uv|vu) > 0.
\end{eqnarray}
This expression is the same as in CIS and USF-TDA, indicating that
if there is no mixing with other types of configurations, the singlet-triplet gap is positive.

Now we examine the effects of mixing other types of configurations. For simplicity, we use the second-order perturbation theory for Eq. \eqref{eq:AX}, such that the correction to $\dST$ is given by
\begin{eqnarray}
\delta\omega^{\mathrm{C}}
\approx
-\sum_{\mu\in \mathrm{C}}
\frac{|\langle \mathrm{C}_\mu|
\widetilde{\mathbf{A}}^{\mathrm{XSF\textrm{-}TDA}}|\mathrm{S}_1\rangle|^2}
{\omega_\mu^{\mathrm{C}}-
\omega_{\textrm{S}_1}^{(0)}
}
\end{eqnarray}
where $\omega_\mu^{\mathrm{C}}$ is the diagonal matrix element of $\widetilde{\mathbf{A}}^{\mathrm{XSF\textrm{-}TDA}}$ for the $\mu$-th excitation within the excitation class $\mathrm{C}\in\{\mathrm{CO(1)},\,\mathrm{OV(1)},\,\mathrm{CV(1)}\}$.
Suppose $\omega_\mu^{\mathrm{C}}>\omega_{\textrm{S}_1}^{(0)}$, which is often the case,
then this correction term lowers $\dST$.

Using the matrix elements for $\widetilde{\mathbf{A}}^{\mathrm{XSF\textrm{-}TDA}}$ given in Ref. \citenum{2025spinadapted}, we can find for the OO(1)-CV(1) type of mixing,
\begin{eqnarray}
\langle \mathrm{C}_\mu|
\widetilde{\mathbf{A}}^{\mathrm{XSF\textrm{-}TDA}}|\mathrm{S}_1\rangle
=
-\frac{\sqrt{3}}{\sqrt{2}}[(au|ui)-(av|vi)],
\end{eqnarray}
with $\mu=ai$, which is greater than the corresponding coupling in USF-TDA by a factor of $\sqrt{3}$.
Therefore, the spin-adapted correction in Eq. \eqref{eq:AX}, arising from the effects of
higher excitations, tends to enhance the OO(1)-CV(1) coupling, resulting in a more negative
$\delta\omega^{\mathrm{CV(1)}}$ under the energy-ordering assumption above. This explains the negative $\dST$ predicted by XSF-TDA 
for the INVEST benchmark by Loos et al,\cite{2023heptazine} where USF-TDA and MRSF-TDA give positive $\dST$ for most functionals. 
In particular, the OO(1)-CV(1) enhancement is missing
in MRSF-TDA as explained in our previous study\cite{2025spinadapted}.
The mechanism of singlet stabilization 
by mixing with the CV(1) type of excitation is similar to the dynamical spin polarization
introduced by Kollmar and Staemmler to explain molecular violations of Hund's rule\cite{kollmar1978hund,koseki1985hund,drwal2023spin}. The closely related term ``double spin polarization'' emphasizes the underlying double-excitation character.\cite{karafiloglou1989dsp} 

Likewise, mixing OV(1) and CO(1) types of excitations can also lower $\dST$, and in this case
XSF-TDA enhances the couplings with the OO(1) type of excitations by roughly a factor of $\sqrt{2}$, compared to the corresponding couplings in USF-TDA. 
Unlike the CV(1) channel, this OO(1)--OV(1)/CO(1) contribution is more naturally associated with relaxation and mixing of the open frontier orbitals with the virtual/core space.

\FloatBarrier
\section{Computational details}\label{sec:comp}
XSF-TDA is implemented in the open-source \textsc{XTDDFT} package\cite{XTDDFTcode} and interoperates with PySCF\cite{2018PySCF,2020Recent} reference calculations. 
A practical XSF-TDA workflow consists of (i) generating a high-spin ROKS reference, (ii) performing XSF-TDA with ALDA0 or MCOL kernel, and (iii) requesting the desired XSF-TDA roots and configuration analysis. The XSF-TDA calculation returns excitation energies and OO(1), CV(1), CO(1), and OV(1) amplitudes, allowing users to diagnose spin recoupling and the DSP-active channels discussed in Sec.~\ref{sec:dsp}. Labels such as XSF-TDA(ALDA0) and XSF-TDA(MCOL) distinguish the kernel used. 

The attenuation factor for the MCOL kernel is calibrated using the ten-molecule triangulene benchmark\cite{2023heptazine} (denoted by Loos-10) with the 6-31G* basis.\cite{Ditchfield1971SelfConsistentMM,hehre1972}
Its transferability is further examined using a 19-molecule open-shell subset of the QUEST database\cite{Loos2025The} (denoted by QUEST-19) with aug-cc-pVTZ.\cite{1989Gaussian}
The four application benchmarks include the NAH159 benchmark set\cite{garner2024,kunze2024},
the 35-molecule MR-TADF benchmark\cite{hall2022,kunze2025} (denoted by MR-TADF-35), the four-molecule experimental heptazine benchmark\cite{0Delayed} (denoted by Heptazine-4), and the 45-molecule INVEST subset derived by Pang et al.\cite{peng2025} (denoted by INVEST-45). To compare with previous DFT/TD-DFT results,
def2-SVP\cite{weigend2005def2} is used for the first three benchmarks, and aug-cc-pVDZ\cite{dunning1989} is used for the last benchmark. The considered XC functionals include
B3LYP\cite{becke1993,lee1988,stephens1994}, PBE0\cite{adamo1999pbe0,perdew1996pbe}, and BHHLYP\cite{becke1993mixing,1988Density,lee1988} throughout the main benchmarks, while BLYP\cite{1988Density,lee1988} is additionally included in the calibration. 

\begin{figure*}[!tp]
    \centering
    \includegraphics[trim=1cm 10cm 1cm 11cm, clip, width=0.95\linewidth]{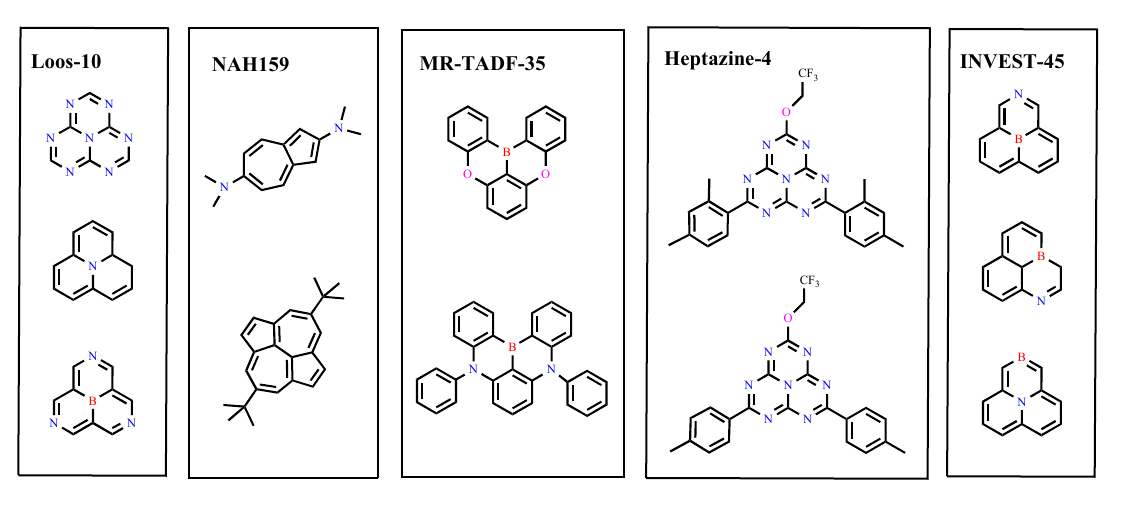}
    \caption{Representative molecules from the Loos-10 calibration set\cite{2023heptazine} and the four application benchmark sets: NAH159,\cite{garner2024,kunze2024} MR-TADF-35,\cite{hall2022,kunze2025} Heptazine-4,\cite{0Delayed} and INVEST-45.\cite{peng2025}
    }
    \label{fig:draw_mo}
\end{figure*}

The four application sets probe distinct aspects of method transferability. 
Some representative molecules are illustrated in Fig. \ref{fig:draw_mo}.
NAH159 contains 159 CC2\cite{CHRISTIANSEN1995409} reference gaps of both positive and negative signs and tests broad chemical coverage.
The MR-TADF-35 benchmark contains 35 experimentally characterized emitters and therefore tests realistic narrow-band chromophores.\cite{hall2022,kunze2025} 
Four experimentally characterized heptazine derivatives provide a stringent near-inversion test,\cite{0Delayed} while the INVEST-45 subset represents a systematic INVEST design space.\cite{peng2025} 
For an error $e_n=\Delta E_{\mathrm{ST},n}^{\mathrm{calc}}-\Delta E_{\mathrm{ST},n}^{\mathrm{ref}}$, we report the mean signed error (MSE), mean absolute error (MAE), root mean squared error (RMSE),
and the signed maximum absolute error (MAXSE), defined as the error with the largest absolute magnitude, retaining its sign:
\begin{align}
\mathrm{MSE}&=N^{-1}\sum_ne_n, \nonumber\\
\mathrm{MAE}&=N^{-1}\sum_n|e_n|,\nonumber\\
\mathrm{RMSE}&=\left(N^{-1}\sum_ne_n^2\right)^{1/2}, \nonumber\\
\mathrm{MAXSE}&=e_{n^*},\;\; n^*=\arg\max_n|e_n|.
\label{eq:statistics}
\end{align}
Sign-classification counts are reported where the reference set contains both positive and negative gaps.

For comparison, singlet-triplet gaps from closed-shell TDA/TD-DFT calculations using restricted Kohn--Sham references and USF-TDA/USF-TD-DFT using unrestricted triplet references are also reported.
ALDA0 and MCOL kernels are compared for USF-TDA and XSF-TDA; the USF-TD-DFT data reported here use the MCOL kernel only\cite{wang2025zero}. All DFT-based calculations were performed with the PySCF infrastructure\cite{2018PySCF,2020Recent} and the \textsc{XTDDFT} implementation.\cite{XTDDFTcode}

\section{Results}\label{sec:result}

\subsection{Calibration of the scaling factor for the MCOL kernel}\label{sec:kernel_results}
\begin{figure*}[!tp]
\centering
\includegraphics[width=1.0\linewidth]{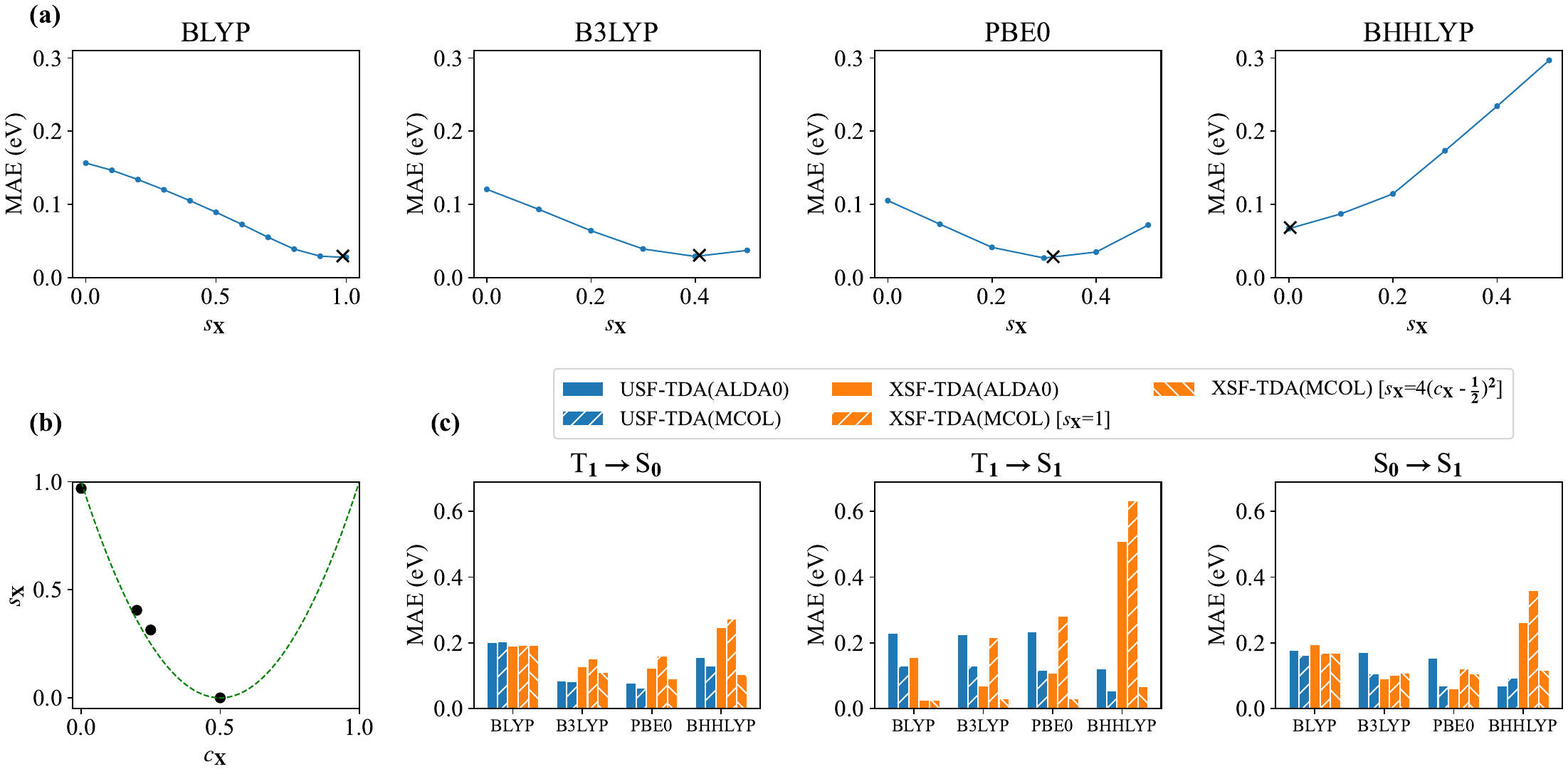}
\caption{Results for the Loos-10 calibration set.
(a) MAE of XSF-TDA(MCOL) with different functionals as a function of $s_\rX$. Cross markers highlight the optimal
$s_\rX$ in [0,1] found by a quadratic fit.
(b) Optimal $s_\rX$ as a function of the amount of Hartree-Fock exchange $c_\rX$. Green dashed line represents the approximate fit
Eq. \eqref{eq:scale}.
(c) Comparison results for excitation energies
(T$_1\rightarrow$ S$_0$ from $\omega_0$, T$_1\rightarrow$ S$_1$ from $\omega_1$, and S$_0\rightarrow $ S$_1$ from $\omega_1-\omega_0$)
obtained from USF-TDA and XSF-TDA.
}
\label{fig:loos-10}
\end{figure*}

The Loos-10 set comprises ten heptazine, cyclazine, and related molecules with inverted singlet–triplet gaps, for which high-level coupled-cluster theoretical best estimates are available.\cite{2023heptazine}
Using this set, we scan $s_\rX$ over $[0,1]$ for BLYP ($c_\rX=0$) and over $[0,0.5]$ for B3LYP ($c_\rX=0.2$), PBE0 ($c_\rX=0.25$), and BHHLYP ($c_\rX=0.5$). The resulting MAEs of $\dST$ as a function of $s_\rX$ are shown in Fig. \ref{fig:loos-10}(a).

We observe that the optimal $s_\rX$ that gives the smallest
MAE decreases as $c_\rX$ increases from 0 to 0.5, see Fig. \ref{fig:loos-10}(b). To fit $s_\rX$ as a function of $c_\rX$ as simple as possible,
we use the obtained data and a condition for the Hartree-Fock limit, that is, $s_\rX=1$ for $c_\rX=1$. Fig. \ref{fig:loos-10}(b) suggests
the use of a simple second-order polynomial
\begin{equation}
s_\rX=4(c_\rX-\frac{1}{2})^2,\label{eq:scale}
\end{equation}
as an approximate parametrization of the fitted trend, while recovering $s_\rX=1$ at the Hartree--Fock limit. Note that for BHHLYP, this turns off
the correction term completely, such that the resulting XSF-TDA equation becomes
similar to the quasi-spin-adapted (QSF-TDA)\cite{chibueze2025spin}, whereas for other values of $c_\rX$, the spin-adapted correction term exists.

Figure \ref{fig:loos-10}(c) compares the MAEs of excitation energies using Eq. \eqref{eq:scale} and without attenuation (i.e., $s_\rX=1$) for XSF-TDA(MCOL).
For Loos-10, the errors of XSF-TDA(MCOL) without attenuation are greater than those of XSF-TDA(ALDA0) in most cases, motivating attenuation of the correction to mitigate
double counting of correlation. The use of Eq. \eqref{eq:scale} generally improves the results for this set. Compared with USF-TDA results, it is found that for transitions that are comparatively insensitive to spin contamination, such as T$_1\rightarrow$ S$_0$ excitations, XSF-TDA retains accuracy comparable to USF-TDA. For BLYP, XSF-TDA is slightly better than USF-TDA, while
USF-TDA is better for B3LYP, PBE0, and BHHLYP.
However, for the spin-sensitive T$_1\rightarrow$ S$_1$ roots, 
XSF-TDA performs better than USF-TDA with both
ALDA0 and MCOL with attenuation, except for BHHLYP.

The Supporting Information also presents the vertical D$_0\rightarrow$ Q$_1$ excitation energies for QUEST-19, defined as $\Delta E_{\mathrm{DQ}}=E(\mathrm{Q}_1)-E(\mathrm{D}_0)=-\omega_0$, where $\omega_0$ is the Q$_1\rightarrow$ D$_0$ response root obtained from the quartet reference (Fig.~S1). As with the T$_1\rightarrow$ S$_0$ transitions in Loos-10, these transitions are comparatively insensitive to spin contamination. Both the effect of attenuation and the accuracy of XSF-TDA relative to USF-TDA depend on the choice of XC functional.
Applications to the benchmark sets discussed below provide a more comprehensive assessment of the method’s transferability.



\subsection{NAH159: broad small- and inverted-gap benchmark}
NAH159 combines aza-aromatic cores with electron-donating and electron-withdrawing substituents, as illustrated in Fig.~\ref{fig:draw_mo}.
The reference vertical singlet--triplet gaps were calculated by Kunze et al.\cite{kunze2024} at the CC2/aug-cc-pVTZ level using the $\omega$B97X-D/def2-TZVP ground-state geometries taken from Garner et al.\cite{garner2024}

\begin{figure*}[!tp]
    \centering
    \includegraphics[width=1.0\linewidth]{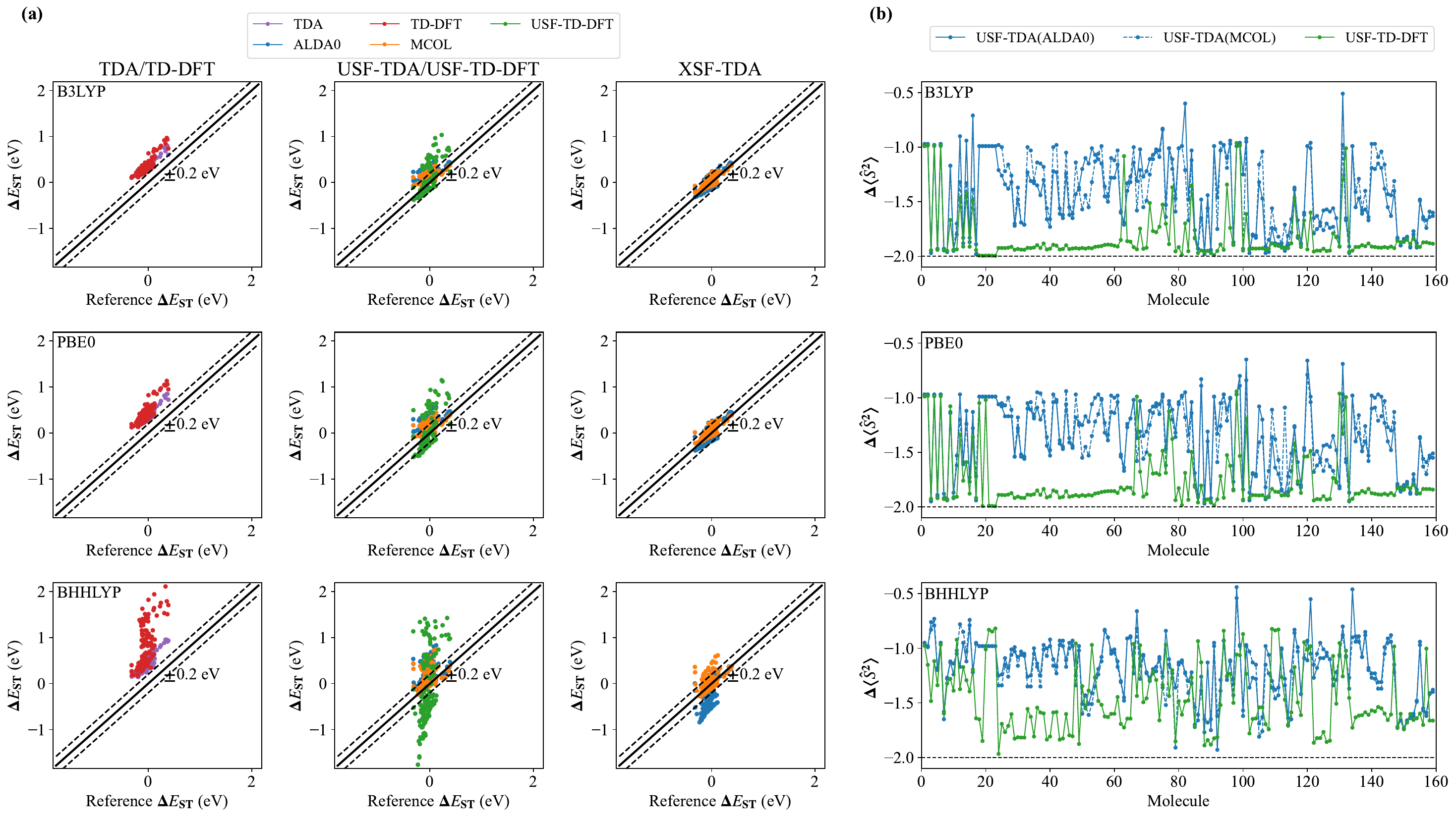} 
    \caption{(a) Calculated $\dST$ (in eV) for the NAH159 benchmark set from different methods with def2-SVP basis against reference CC2 data.
    (b) $\dSpinSquare$ for the S$_1$ state from USF-TDA(ALDA0), USF-TDA(MCOL) and USF-TD-DFT.
    }
    \label{fig:nah159}
\end{figure*}

\begin{table*}[!tp]
    \centering
    \resizebox{\linewidth}{!}{
    \begin{tabular}{cccccccc}
    \hline\hline
        Functional & Method & MSE & MAE & RMSE & MAXSE & \makecell{Correct Sign\\(Positive)} & \makecell{Correct Sign\\(Negative)}   \\ \hline
        B3LYP & TDA & 0.356 & 0.356 & 0.361 & 0.480 & 40/40 & 0/119  \\ 
        ~ & TD-DFT & 0.379 & 0.379 & 0.387 & 0.630 & 40/40 & 0/119  \\ 
        ~ & USF-TDA(ALDA0) & 0.282 & 0.282 & 0.302 & 0.551 & 40/40 & 0/119   \\ 
        ~ & USF-TDA(MCOL) & 0.178 & 0.181 & 0.198 & 0.437 & 40/40 & 18/119 \\ 
        ~ & USF-TD-DFT(MCOL) & 0.048 & 0.149 & 0.232 & 0.863 & 30/40 & 96/119 \\
        ~ & XSF-TDA(ALDA0) & -0.010 & 0.067 & 0.087 & 0.233 & 23/40 & 103/119 \\ 
        ~ & XSF-TDA(MCOL) & 0.036 & 0.057 & 0.073 & 0.249 & 30/40 & 98/119   \\ 
        \hline
        PBE0 & TDA & 0.384 & 0.384 & 0.390 & 0.537 & 40/40 & 0/119  \\ 
        ~ & TD-DFT & 0.446 & 0.446 & 0.459 & 0.818 & 40/40  & 0/119  \\ 
        ~ & USF-TDA(ALDA0) & 0.315 & 0.315 & 0.338 & 0.621 & 40/40  & 0/119 \\ 
        ~ & USF-TDA(MCOL) & 0.183 & 0.185 & 0.206 & 0.491 & 40/40 & 29/119 \\ 
        ~ & USF-TD-DFT(MCOL) & 0.013 & 0.230 & 0.305 & 0.927 & 24/40 & 100/119 \\
        ~ & XSF-TDA(ALDA0) & -0.027 & 0.087 & 0.107 & 0.284 & 22/40  & 104/119  \\ 
        ~ & XSF-TDA(MCOL) & 0.050  & 0.067 & 0.089 & 0.337 & 31/40 & 90/119  \\ 
        \hline
        BHHLYP & TDA & 0.434 & 0.434 & 0.444 & 0.681  & 40/40  & 0/119  \\ 
        ~ & TD-DFT & 0.818 & 0.818 & 0.905 &1.825  & 40/40  & 0/119  \\ 
        ~ & USF-TDA(ALDA0) & 0.251 & 0.255 &0.317  & 0.857 & 40/40  &24/119  \\ 
        ~ & USF-TDA(MCOL) &0.169  &0.185  &0.251  &0.760  &36/40 &55/119  \\ 
        ~ & USF-TD-DFT(MCOL) & -0.086 & 0.610 & 0.703 & -1.531 & 24/40 & 88/119 \\
        ~ & XSF-TDA(ALDA0) &-0.301  &0.330  &0.369  &-0.609  &21/40   &114/119   \\ 
        ~ & XSF-TDA(MCOL) &0.063   &0.110  &0.166  &0.684  &30/40  &83/119   \\
        \hline
        \multicolumn{2}{c}{$\Delta$PBE0@HL\cite{kunze2024}} & 0.026 & 0.116 & 0.285 & 2.279 & 23/40 & 101/119 \\
        \multicolumn{2}{c}{$\Delta$PBE0@2$\times$2\cite{kunze2024}} & 0.002 & 0.053 & 0.069 & 0.274 & 25/40 & 107/119 \\
        \multicolumn{2}{c}{QE-B3LYP@(N+1)\cite{li2026}} & 0.031 & 0.090 & 0.164 & 1.070 & 30/40 & 96/119 \\
        \multicolumn{2}{c}{QE-B3LYP@(N-1)\cite{li2026}} & -0.004 & 0.108 & 0.164 & 0.895 & 27/40 & 109/119 \\
        \hline\hline
    \end{tabular} }
    \caption{Statistical analysis of $\dST$ (in eV) obtained from different methods and comparison with previous results for the NAH159 benchmark set.}
    \label{tab:nah159}
\end{table*}

As shown in Fig.~\ref{fig:nah159}(a),
conventional TDA and TD-DFT overestimate most gaps by more than 0.2 eV, regardless of the XC functionals in use. USF-TDA reduces $\dST$, but still retains a strong positive bias.
Compared with USF-TDA(MCOL), USF-TD-DFT further lowers $\dST$ and reduces the magnitude of the MSE. The sign-classification counts in Table~\ref{tab:nah159} show improved recovery of negative gaps, at the expense of fewer correctly classified positive gaps. However, the larger RMSE and $|\mathrm{MAXSE}|$ indicate larger errors in the predicted gaps. 
Following Ref.~\citenum{li2011spin}, we monitor spin contamination in USF-TDA/USF-TD-DFT through
the spin-square change $\dSpinSquare=\langle\hat S^2\rangle_{\mathrm{ex}}-\langle\hat S^2\rangle_{\mathrm{ref}}$; for a spin-flip-down target with $S_f=S_i-1$, the ideal change is $-2S_i$ (equal to $-2$ for a triplet reference). As shown in Fig.~\ref{fig:nah159}(b), 
$\Delta\langle\hat S^2\rangle$ for the S$_1$ state in USF-TDA deviates from the ideal values significantly for most of molecules in NAH159.
USF-TD-DFT improves the spin diagnostic significantly for B3LYP and PBE0, but only slightly for BHHLYP. Relative to USF-TDA(MCOL), XSF-TDA(MCOL) reduces the positive bias in $\dST$ and improves agreement with the CC2 reference for all three XC functionals. XSF-TDA(ALDA0) also improves the accuracy for B3LYP and PBE0, but overcorrects the gaps for BHHLYP, increasing the MAE from 0.255 to 0.330 eV relative to USF-TDA(ALDA0). In particular, the MAE of XSF-TDA(ALDA0) and XSF-TDA(MCOL) with the B3LYP functional is only 0.067 eV and 0.057 eV, respectively.

Table~\ref{tab:nah159} also compares the results from previous studies using $\Delta$PBE0\cite{kunze2024} with the def2-SVP basis set
and QE-B3LYP\cite{li2026} with the def2-TZVP basis set.
We find that XSF-TDA with B3LYP outperforms QE-B3LYP with both the $(N+1)$ and $(N-1)$ formulations\cite{li2026},
with smaller MAE and RMSE, while reducing MAXSE by a factor of 4-5.
The performance of XSF-TDA is also much better than $\Delta$PBE with a HOMO--LUMO-only window ($\Delta$PBE0@HL), and is comparable to $\Delta$PBE0 with a 2$\times$2 orbital window ($\Delta$PBE0@2$\times$2)\cite{kunze2024}. 
Therefore, XSF-TDA reaches accuracy competitive with these state-specific approaches, while retaining a response-state description that enables a deeper mechanistic analysis carried out in Sec. \ref{sec:Mechanism}.

\subsection{MR-TADF-35: experimentally characterized MR-TADF emitters}
For MR-TADF-35, the experimental reference gaps were taken from the fluorescence and phosphorescence peak energies recompiled by Kunze et al.\cite{kunze2025} These emission-derived gaps are compared with calculated vertical gaps at ground-state geometries following their benchmark protocol, although the two quantities are not strictly equivalent.

\begin{figure*}[!tp]
    \centering
    \includegraphics[width=1.0\linewidth]{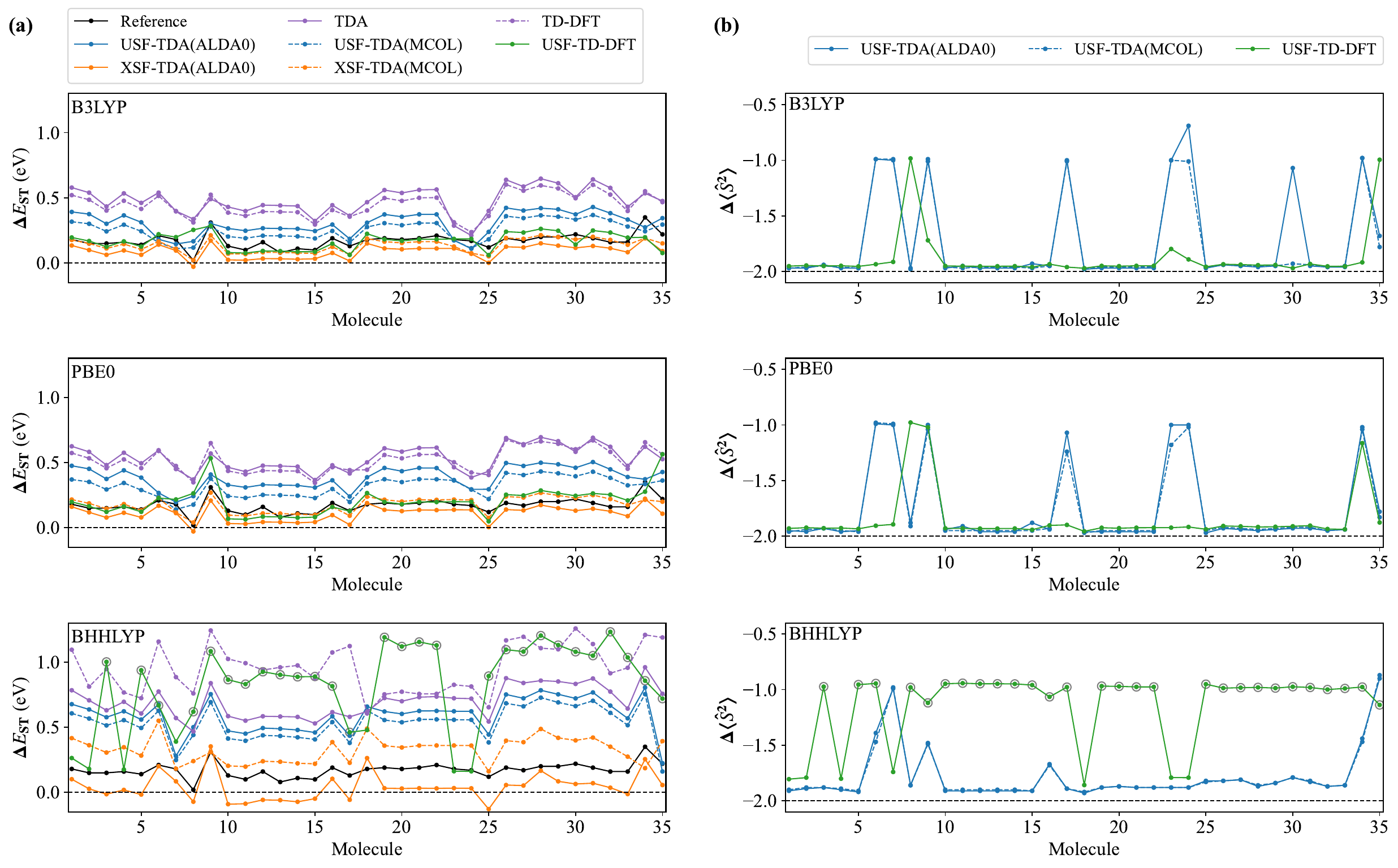} 
    \caption{(a) Calculated $\dST$ (in eV) for the MR-TADF-35 benchmark set from different methods using def2-SVP basis against experimental reference gaps\cite{kunze2025}.
    (b) $\dSpinSquare$ for the S$_1$ state from USF-TDA(ALDA0), USF-TDA(MCOL), and USF-TD-DFT. Open circles in the USF-TD-DFT/BHHLYP results indicate that the corresponding excitation energy is complex.
    }
    \label{fig:mewes}
\end{figure*}

\begin{table*}[!tp]
    \centering
    \begin{tabular}{cccccc}
    \hline\hline
        Functional & Method & MSE & MAE & RMSE & MAXSE   \\ \hline
        B3LYP & TDA & 0.310 & 0.310 & 0.324 & 0.451 \\
        & TD-DFT & 0.277 & 0.277 & 0.287 & 0.412\\
        ~ & USF-TDA(ALDA0) & 0.134  & 0.146  & 0.160  & 0.240    \\ 
        ~ & USF-TDA(MCOL) & 0.083  & 0.102  & 0.111  & 0.178  \\ 
        ~ & USF-TD-DFT(MCOL) & -0.002 & 0.046 & 0.066 & 0.234 \\
        ~ & XSF-TDA(ALDA0) & -0.081  & 0.081  & 0.087  & -0.161   \\ 
        ~ & XSF-TDA(MCOL) & -0.035  & 0.039  & 0.052  & -0.159  \\ 
        \hline
        PBE0 & TDA & 0.364 & 0.364 & 0.372 & 0.500\\
        & TD-DFT & 0.347 & 0.347 & 0.353 & 0.486 \\
        ~ & USF-TDA(ALDA0) & 0.214  & 0.214  & 0.228  & 0.314   \\ 
        ~ & USF-TDA(MCOL) & 0.146  & 0.149  & 0.160  & 0.239    \\ 
        ~ & USF-TD-DFT(MCOL) & 0.032 & 0.059 & 0.093 & 0.343 \\
        ~ & XSF-TDA(ALDA0) & -0.063  & 0.063  & 0.070  & -0.128    \\ 
        ~ & XSF-TDA(MCOL) & 0.007  & 0.035  & 0.043  & -0.136  \\ 
        \hline
        BHHLYP & TDA & 0.527 & 0.527 & 0.533 & 0.686\\
        & TD-DFT & 0.786  &0.786  & 0.803 &1.037  \\
        ~ & USF-TDA(ALDA0) &0.421   &0.421   &0.437   & 0.583   \\ 
        ~ & USF-TDA(MCOL) &0.363   &0.366   &0.380   &0.527     \\ 
        ~ & USF-TD-DFT(MCOL) & 0.648 & 0.649 &0.725  &1.071  \\
        ~ & XSF-TDA(ALDA0) &-0.129   &0.136   &0.145   & -0.248    \\ 
        ~ & XSF-TDA(MCOL) & 0.156  &0.165   &0.181   & 0.340  \\
        \hline
        \multicolumn{2}{c}{$\Delta$UKS/LC-$\omega$PBE\cite{kunze2025}} & 0.011 & 0.054 & 0.073 & 0.248 \\
        \multicolumn{2}{c}{$\Delta$UKS/FX175-$\omega$PBE\cite{kunze2025}} & -0.015 & 0.029 & 0.047 & -0.193\\
        \multicolumn{2}{c}{QE-B3LYP@(N+1)\cite{li2026}} & -0.033 & 0.045 & 0.059 & -0.185\\
        \multicolumn{2}{c}{QE-B3LYP@(N-1)\cite{li2026}} & -0.050 & 0.058 & 0.079 & -0.207\\
        \multicolumn{2}{c}{SCS-CC2/cc-pVDZ\cite{hall2022}} & -0.006	& 0.040	& 0.049 & -0.110 \\
        \hline\hline
    \end{tabular}
    \caption{
    Statistical analysis of $\dST$ (in eV)  obtained from different methods and comparison with previous results for the MR-TADF-35 benchmark set.
    }
    \label{tab:meses35}
\end{table*}

Unlike NAH159, all $\dST$ in MR-TADF-35 are small, but positive values.
Figure \ref{fig:mewes}(a) shows that TDA and TD-DFT overestimate the experimental gaps with their MAEs exceed 0.27 eV, see Table \ref{tab:meses35}. In this benchmark set, most USF-TDA and USF-TD-DFT results have $\Delta\langle S^2\rangle$ close to $-2$,
see Fig. \ref{fig:mewes}(b), indicating that spin contamination is less severe here than in NAH159. Nevertheless, USF-TDA remains positively biased. 
USF-TD-DFT improves the agreement with the reference values significantly for B3LYP and PBE0, with MAEs being 0.046 and 0.059 eV, respectively.
However, USF-TD-DFT/BHHLYP excitation energies are complex for some molecules,
see the points highlighted by open circles in Fig. \ref{fig:mewes}. In this case,
we take the real part of the excitation energy as $\dST$.
XSF-TDA improves USF-TDA for both ALDA0 and MCOL kernels, demonstrating that
the benefit is not restricted to strongly spin-contaminated states.
We notice that XSF-TDA with ALDA0 tends to overcorrect $\dST$, as indicated by the negative MSE. The additional attenuation in MCOL mitigates this problem, leading to smaller MAEs of 0.039 and 0.035 for B3LYP and PBE0, respectively.

For comparison, Table \ref{tab:meses35} also lists previous results obtained using state-specific methods. $\Delta$UKS with LC-$\omega$PBE and FX175-$\omega$PBE gives MAEs of 0.054 and 0.029 eV\cite{kunze2025}, respectively, while QE-B3LYP gives 0.045 and 0.058 eV in the $(N+1)$ and $(N-1)$ formulations\cite{li2026}, respectively. The SCS-CC2/cc-pVDZ results\cite{hall2022} have an MAE of 0.040 eV against the reference values on this emitter benchmark. Therefore, XSF-TDA/PBE0 is comparable to the best reported state-specific DFT and SCS-CC2 results.



\subsection{Heptazine-4: experimentally characterized heptazine emitters}
Aizawa and co-workers screened 34596 heptazine derivatives generated from 186 substituents and experimentally characterized \ce{HzTFEX2}, \ce{HzTFEP2}, \ce{HzTFET2}, and \ce{HzPipX2}.\cite{0Delayed} Table~\ref{tab:miyajima} compares their experimental gaps with the present calculations and previous QE-DFT, MRSF-TDA, and wave-function results.

\begin{table*}[!tp]
   	\centering
   	\begin{tabular}{cccccc}
   		\hline\hline
   		Functional & Method & \ce{HzTFEX2} & \ce{HzTFEP2} & \ce{HzTFET2} & \ce{HzPipX2} \\
   		\hline
   	
        B3LYP &TDA & 0.227 & 0.262 & 0.258 & 0.280  \\
        & TD-DFT & 0.222 & 0.262 & 0.261 & 0.281 \\
   		& USF-TDA(ALDA0) &  0.235 & 0.279 & 0.277 & 0.293 \\
        & & (-1.25) & (-1.29) & (-1.32) & (-1.47)\\
   		& USF-TDA(MCOL) & 0.139 & 0.168 & 0.168 & 0.183 \\
        & & (-1.16) & (-1.26) & (-1.29) & (-1.52) \\
        & USF-TD-DFT(MCOL) & -0.185 & -0.149 & -0.138 & -0.095 \\
        & & (-1.90) & (-1.87) & (-1.87) & (-1.87) \\
   		& XSF-TDA(ALDA0) & -0.213 & -0.202 & -0.198 & -0.169  \\
   		& XSF-TDA(MCOL) & -0.112 & -0.102 & -0.099 & -0.071\\
        \hline
        PBE0 & TDA & 0.246 & 0.282 & 0.279 & 0.305 \\
        & TD-DFT &0.248 & 0.294 & 0.297 & 0.318 \\ 
   		& USF-TDA(ALDA0) &  0.288 & 0.337 & 0.335 & 0.342 \\
        & & (-1.12) & (-1.13) & (-1.14) & (-1.39)\\
   		& USF-TDA(MCOL) & 0.174 & 0.208 & 0.209 & 0.217 \\
        & & (-1.01) & (-1.09) & (-1.11) & (-1.43) \\
        & USF-TD-DFT(MCOL) & -0.281 & -0.226 & 0.460 & -0.158 \\
        & & (-1.85) & (-1.82) & (-0.99) & (-1.83) \\
   		& XSF-TDA(ALDA0) & -0.256 & -0.243 & -0.237 & -0.203 \\
   		& XSF-TDA(MCOL) & -0.104 & -0.092 & -0.088 & -0.053 \\
        \hline
        BHHLYP & TDA &0.283  & 0.312 &0.327  &0.357  \\
        & TD-DFT & 0.754 &0.684 &0.750 & 0.687 \\ 
   		& USF-TDA(ALDA0) &0.098  &0.178 &0.194 & 0.475 \\
        & & (-1.10) & (-1.06) & (-1.05) & (-0.95)\\
   		& USF-TDA(MCOL) &0.023  &0.093  & 0.109 &0.363 \\
        & & (-1.14) & (-1.07) & (-1.05) & (-0.98) \\
        & USF-TD-DFT(MCOL) &0.918  &1.193  & 1.065 &1.057  \\
        & & (-0.97) & (-0.98) & (-0.98) & (-0.92) \\
   		& XSF-TDA(ALDA0) &-0.640  &-0.611  &-0.599  & -0.523 \\
   		& XSF-TDA(MCOL) &-0.111 &-0.096 &-0.090  &-0.002  \\
        \hline
        \multicolumn{2}{c}{QE-B3LYP@(N+1)\cite{li2026}} &  -0.084 &  -0.042 & -0.046 & -0.016 \\
        \multicolumn{2}{c}{QE-B3LYP@(N-1)\cite{li2026}} & -0.116 & -0.098 & -0.050 & -0.041 \\
        \multicolumn{2}{c}{MRSF-TDA/DTCAM-STG\cite{cheol2025mr}} & -0.055 & -0.076 & -0.069 & 0.059 \\
        \multicolumn{2}{c}{ADC(2)\cite{0Delayed}} & -0.034 & -0.041 &  -0.082 & -0.012    \\
        \multicolumn{2}{c}{EOM-CCSD\cite{0Delayed}} & -0.012 & -0.033 & -0.045 &  0.010    \\
        \multicolumn{2}{c}{CASPT2(12,12)\cite{0Delayed}} & -0.184 &  -0.355 &  -0.264 & -0.171 \\
        \multicolumn{2}{c}{Expt.\cite{0Delayed}} & -0.011 & -0.014 & -0.013 & 0.052\\
   		\hline\hline
   	\end{tabular}
    \caption{Comparison of calculated $\dST$ (in eV) with different methods and  previously reported results for the Heptazine-4 benchmark set. Data in parentheses indicate the values of $\dSpinSquare$ for the S$_1$ state obtained in USF-TDA or USF-TD-DFT calculations.}
   	\label{tab:miyajima}
\end{table*}
 
The geometries used here were taken from Ref.~\citenum{li2026}, which were optimized at B3LYP-D3/def2-SVP.\cite{grimme2010}  
The calculated $\dST$ with different methods are summarized in Table \ref{tab:miyajima}.
TDA, TD-DFT, and USF-TDA yield positive gaps throughout and miss the three experimentally negative cases. XSF-TDA predicts the three negative gaps with the correct sign, but also predicts a slightly negative gap for \ce{HzPipX2}, whose experimental value is positive.
For comparison, QE-B3LYP/def2-TZVP in the $(N+1)$ and $(N-1)$ formulations also predict all negative gaps. 

Parameter-optimized MRSF-TDA/DTCAM-STG\cite{cheol2025mr} yields the correct sign for \ce{HzPipX2}. 
The published wave-function results were evaluated at MP2-optimized triplet geometries,\cite{0Delayed} so their comparison with the present results also includes differences in the molecular geometries. ADC(2), and CASPT2(12,12) predict all negative gaps, while 
EOM-CCSD gives the correct sign \ce{HzPipX2}. 

We should emphasize these kinetically derived experimental gaps\cite{0Delayed} are more closely related to adiabatic gaps. Their comparison with vertical gaps should therefore be interpreted with this distinction in mind. This small benchmark illustrates both the utility of XSF-TDA near singlet–triplet inversion and the challenge of resolving energy differences of only a few hundredths of an electronvolt. Analytic gradients are under development to enable excited-state geometry optimizations and a more direct comparison of calculated adiabatic gaps with experiment in future.

\subsection{INVEST-45: designed INVEST candidates}

We analyze a derived 45-molecule INVEST subset by Pang et al.,\cite{peng2025} focusing on molecules with negative energy gaps. The reference energy gaps are computed in this work at the CC2/aug-cc-pVDZ level using the Q-Chem package,\cite{qchem} with molecular geometries taken from Pang et al.\cite{qchem,peng2025} optimized at the B3LYP/cc-pVDZ level.

\begin{figure*}[!tp]
    \centering
    \includegraphics[width=1.0\linewidth]{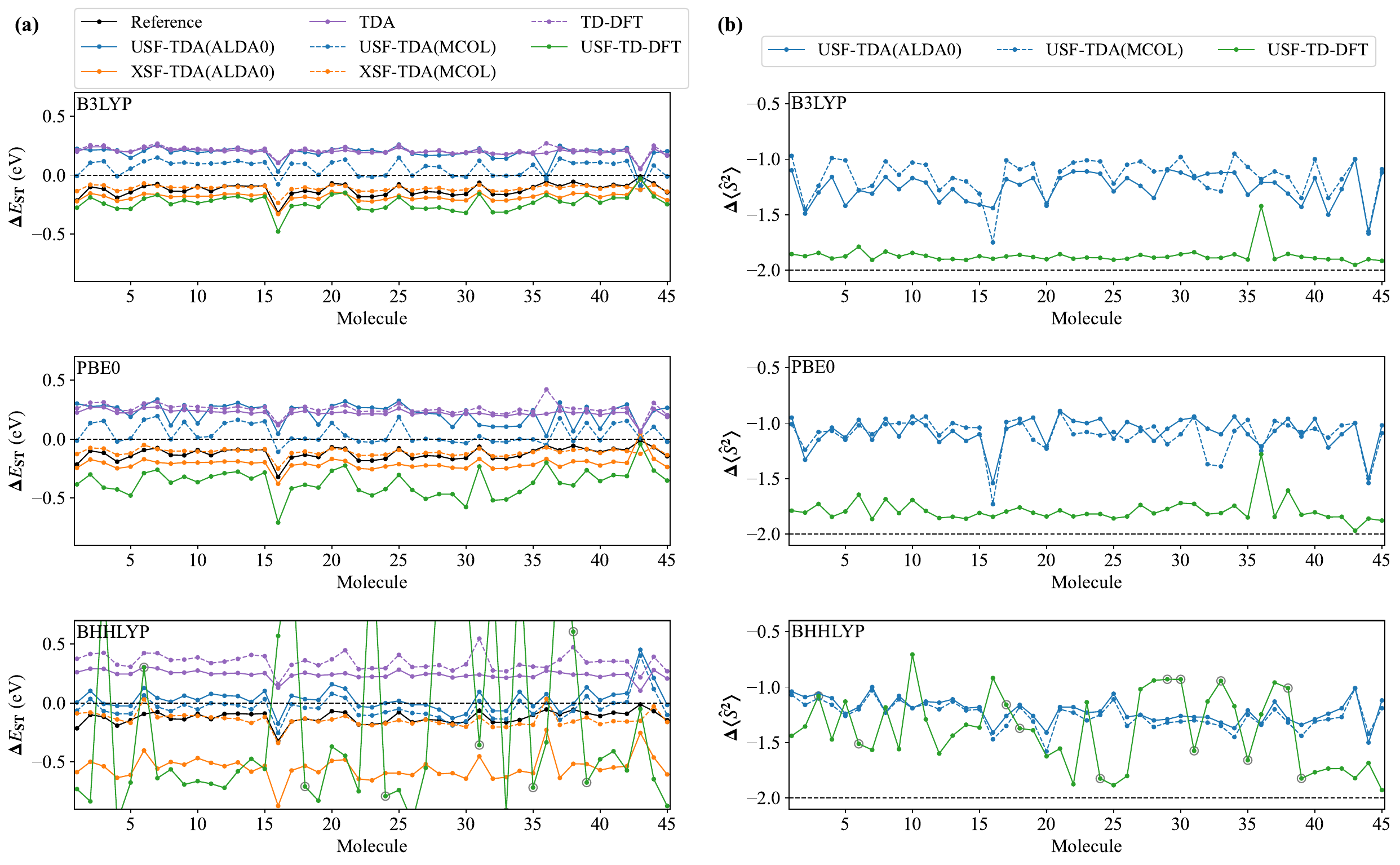} 
    \caption{(a) Calculated $\dST$ (in eV) for the INVEST-45 benchmark set from different methods using aug-cc-pVDZ basis against CC2 reference gaps.
    (b) $\dSpinSquare$ for the S$_1$ state from USF-TDA(ALDA0), USF-TDA(MCOL), and USF-TD-DFT. Open circles in the USF-TD-DFT/BHHLYP results indicate that the corresponding excitation energy is complex.
    }
    \label{fig:invest45}
\end{figure*}

\begin{table*}[!tp]
    \centering
    \begin{tabular}{cccccc}
    \hline\hline
        Functional & Method & MSE & MAE & RMSE & MAXSE   \\ 
        \hline
        B3LYP & TDA & 0.321 & 0.321 & 0.325 & 0.424 \\
        ~ & TD-DFT & 0.333 & 0.333 & 0.337 & 0.428 \\
        ~ & USF-TDA(ALDA0) & 0.311 & 0.314 & 0.321 & 0.439 \\
        ~ & USF-TDA(MCOL) & 0.184 & 0.189 & 0.194 & 0.254 \\
        ~ & USF-TD-DFT(MCOL) & -0.112 & 0.112 & 0.116 & -0.189 \\
        ~ & XSF-TDA(ALDA0) & -0.060 & 0.060 & 0.064 & -0.099 \\
        ~ & XSF-TDA(MCOL) & 0.013 & 0.028 & 0.035 & -0.112 \\
        \hline
        PBE0 & TDA & 0.344 & 0.344 & 0.349 & 0.440 \\
        ~ & TD-DFT & 0.379 & 0.379 & 0.383 & 0.477 \\
        ~ & USF-TDA(ALDA0) & 0.343 & 0.343 & 0.356 & 0.517 \\
        ~ & USF-TDA(MCOL) & 0.170 & 0.170 & 0.182 & 0.284 \\
        ~ & USF-TD-DFT(MCOL) & -0.245 & 0.245 & 0.256 & -0.416 \\
        ~ & XSF-TDA(ALDA0) & -0.086 & 0.089 & 0.092 & -0.140 \\
        ~ & XSF-TDA(MCOL) & 0.013 & 0.026 & 0.034 & -0.114 \\
        \hline
        BHHLYP & TDA & 0.365 & 0.365 & 0.369 & 0.477 \\
        ~ & TD-DFT & 0.471 & 0.471 & 0.475 & 0.611 \\
        ~ & USF-TDA(ALDA0) & 0.156 & 0.156 & 0.172 & 0.464 \\
        ~ & USF-TDA(MCOL) & 0.085 & 0.088 & 0.111 & 0.414 \\
        ~ & USF-TD-DFT(MCOL) & -0.063 & 0.706 & 0.803 & 1.620 \\
        ~ & XSF-TDA(ALDA0) & -0.426 & 0.426 & 0.432 & -0.552 \\
        ~ & XSF-TDA(MCOL) & -0.019 & 0.043 & 0.056 & -0.139 \\
        \hline\hline
    \end{tabular}
    \caption{Statistical analysis of $\dST$ (in eV) obtained from different methods for the INVEST-45 subset.}
    \label{tab:invest45}
\end{table*}

Figure~\ref{fig:invest45} summarizes the obtained results from different methods. As for NAH159, TDA and TD-DFT predict exclusively positive gaps for all considered functionals. USF-TDA lowers the gaps but remains too positive and exhibits substantial spin contamination, see Fig.~\ref{fig:invest45}(b).
USF-TD-DFT significantly lowers $\dST$, and tends to over-stabilize the singlet. The magnitude
of overstablization is sensitive the amount of Hartree-Fock exchange. For BHHLYP,
complex roots appear for some systems, as marked by gray circles.
XSF-TDA(ALDA0) improves USF-TD-DFT, but also over-stabilizes the singlet,
while XSF-TDA(MCOL) provides the best balance. As shown in Table~\ref{tab:invest45}, the MAEs are 0.028, 0.026, and 0.043 eV for B3LYP, PBE0, and BHHLYP, respectively.

\subsection{Cross-dataset performance}
\begin{figure*}[!tp]
    \centering
    \includegraphics[width=1.0\linewidth]{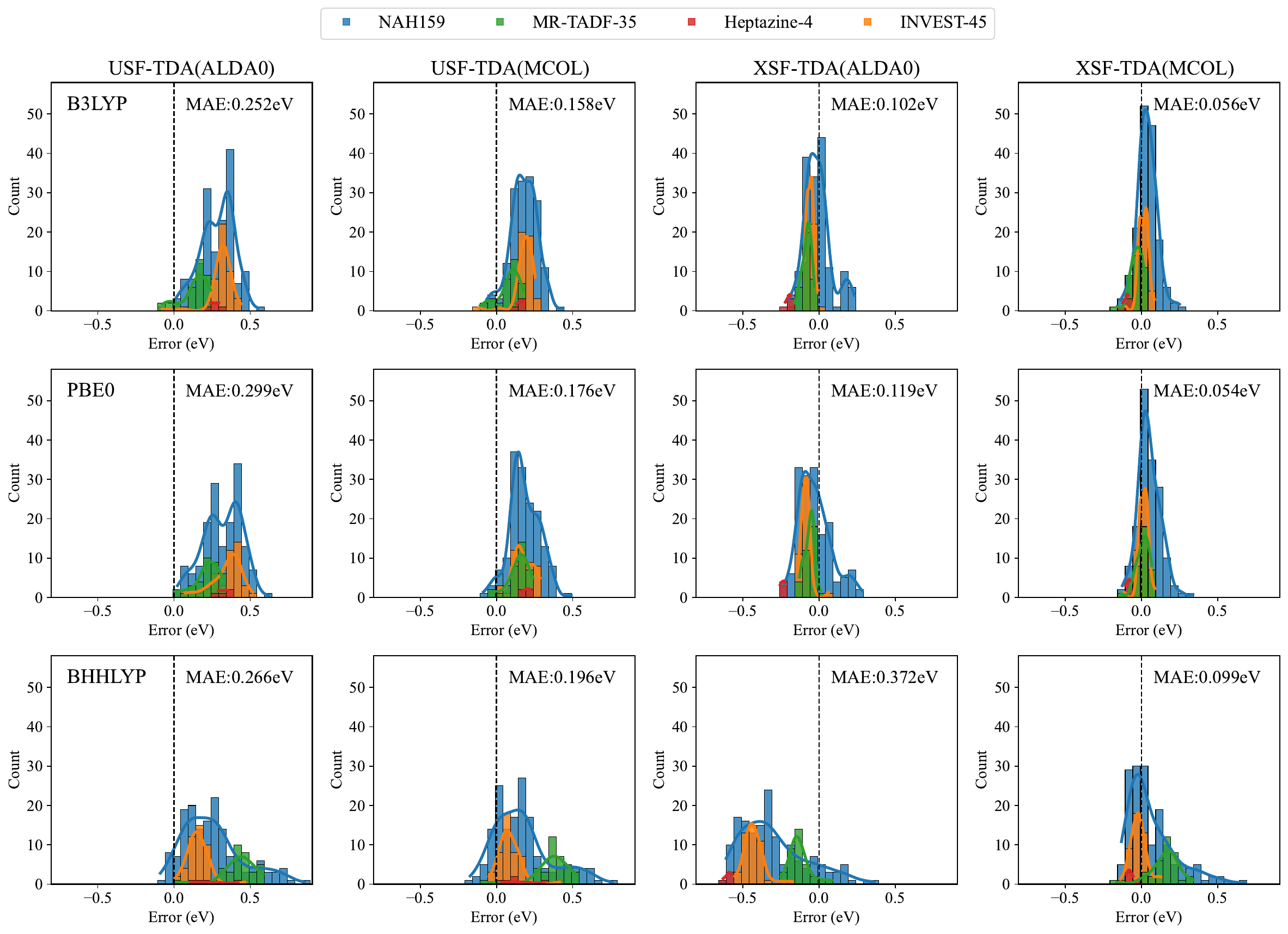}
    \caption{Error histograms for computed $\dST$ (in eV) across the NAH159, MR-TADF-35, Heptazine-4, and INVEST-45 sets. Columns: USF-TDA(ALDA0), USF-TDA(MCOL), XSF-TDA(ALDA0), and XSF-TDA(MCOL); Rows: B3LYP, PBE0, and BHHLYP. The MAE printed in each panel is the equal-set-weighted macro-MAE in Eq.~\ref{eq:macro_mae}, not the MAE of the pooled 243 entries.}
    \label{fig:all_in_bar}
\end{figure*}

Figure~\ref{fig:all_in_bar} displays the error histograms of computed $\dST$ across all the four benchmark sets for USF-TDA(ALDA0), USF-TDA(MCOL), XSF-TDA(ALDA0), and XSF-TDA(MCOL) with different XC functionals, which reveal whether improvement is systematic or confined to a single benchmark
Because the four application sets contain very different numbers of molecules with different electronic structures, the MAE printed in each panel of Fig.~\ref{fig:all_in_bar} is the equal-set-weighted macro-average
\begin{equation}
\mathrm{MAE}_{\mathrm{macro}}
=\frac{1}{4}\sum_{d=1}^{4}\mathrm{MAE}_{d},
\label{eq:macro_mae}
\end{equation}
where $d$ labels NAH159, MR-TADF-35, Heptazine-4, or INVEST-45. 

The combined error distributions make the method-level trend clear. Both USF-TDA variants remain predominantly positive, whereas XSF-TDA(ALDA0) shifts the distribution too far toward negative errors, in particular, for XC functionals with larger amount of Hartree-Fock exchange. 
XSF-TDA(MCOL) centers the B3LYP and PBE0 distributions most effectively, reducing the corresponding macro-MAEs to 0.056 and 0.054~eV. For a given XC functional, the MAE of XSF-TDA(MCOL) is found to be smaller than that of USF-TDA(MCOL), demonstrating a very consistent improvement.

\subsection{Mechanistic analysis}\label{sec:Mechanism}
\begin{figure*}[!htbp]
    \centering
    \includegraphics[width=1.0\linewidth]{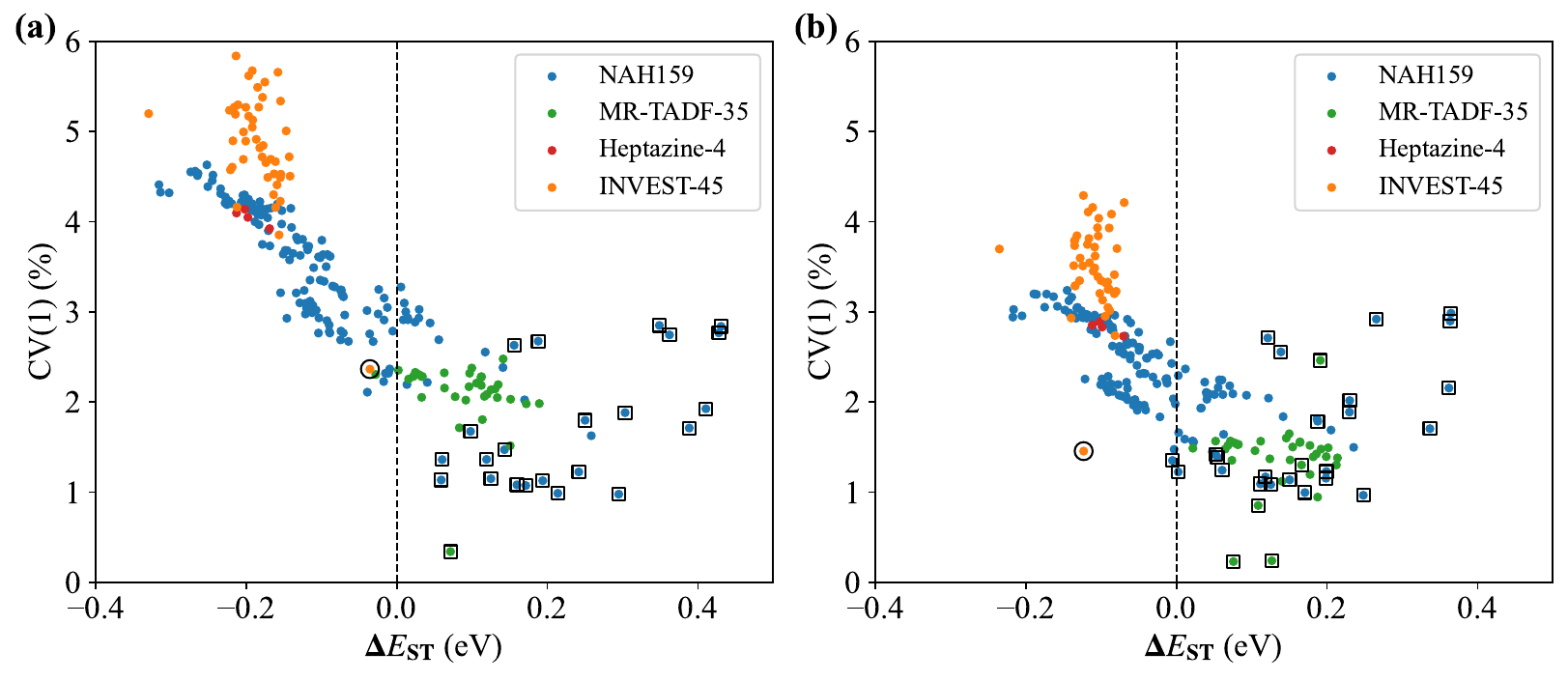}
    \caption{Computed CV(1) weights in the S$_1$ state versus $\dST$ at the
    XSF-TDA/B3LYP level with (a) ALDA0 and (b) MCOL kernels. 
    Open squares mark the S$_1$ states with the OO(1) contribution below 50\%, 
    while the open circle highlights an outlier from INVEST-45, whose negative $\dST$, low CV(1) weight, and appreciable OV(1) character suggest a possible contribution from OO(1)--OV(1) coupling.}
    \label{fig:cv}
\end{figure*}

An advantage of XSF-TDA over state-specific $\Delta$DFT or QE-DFT is that the mechanism of
stabilization can be analyzed from the amplitudes $\tilde{\mathbf{X}}_\lambda$ from Eq. \eqref{eq:AX}.
Here, we focus on the weights of different classes of excitations for the $\lambda$-th eigenvector
\begin{equation}
w_{\mathrm{C}}
=\sum_{pq\in \mathrm{C}}\left|\tilde{X}_{pq,\lambda}^{\mathrm{C}}\right|^2,
\label{eq:cv_weight}
\end{equation}
where the class $\mathrm{C}\in\{\mathrm{OO(1),\,CO(1),\,OV(1),\,CV(1)}\}$.
We first analyze the role of DSP for stabilizing the S$_1$ state.
Figure~\ref{fig:cv} compares the computed CV(1) weights in the S$_1$ state versus $\dST$ for all four benchmark sets at the XSF-TDA/B3LYP level with ALDA0 and MCOL kernels. 
The CV(1) weights are generally smaller with the attenuated MCOL scheme than with ALDA0, which is consistent with the mitigation of excessive singlet stabilization. 
Open squares mark the S$_1$ states with the OO(1) contribution below 50\%. After excluding these points, the combined data suggest a qualitative tendency for larger CV(1) weights to be associated with smaller $\dST$, most clearly within NAH159. Although the strength of this association varies across the benchmark sets, the observed tendency is broadly consistent with the OO(1)--CV(1) stabilization discussed in Sec. \ref{sec:dsp}.

A clear outlier from INVEST-45 is marked by open circles, whose CV(1) weights are much smaller than those for molecules from the same set. For this molecule (no. 43 in INVEST-45
illustrated in the last figure in Fig. \ref{fig:draw_mo}), $\dST$ are -0.036 and -0.123 eV
for XSF-TDA/B3LYP with ALDA0 and MCOL kernels, respectively.
CC2 also gives a small negative gap being -0.012 eV.
The OV(1) character in S$_1$ is found to be 6.6\% with MCOL, 
suggesting substantial mixing between OO(1) and OV(1) configurations beyond the standard DSP channel. To the best of our knowledge, this singlet-stabilization channel has not been emphasized in prior MR-TADF/INVEST studies. Therefore,
analysis of the XSF-TDA response amplitudes can help to identify candidate stabilization channels for further investigation.

\section{Conclusion}\label{sec:conclusion}
In this work, we demonstrate that XSF-TDA offers a practical computational tool for the small correlation-controlled singlet--triplet gaps encountered in MR-TADF and INVEST emitters. Across four chemically distinct application sets, it substantially reduces the strong positive bias of TDA and USF-TDA.
The method also returns a chemically readable explanation. The CV(1) configurations in XSF-TDA
are spin-adapted combinations of the nominal core-to-virtual spin-flip-down determinant and the companion determinants required for spin-adaptation. 
Mixing these configurations with the OO(1) singlet provides the main DSP-like route to a smaller $\dST$. 
XSF-TDA also identifies an INVEST molecule with appreciable OV(1) character, suggesting a possible additional contribution from OO(1)--OV(1) mixing to singlet stabilization. 

An open-source implementation\cite{XTDDFTcode} of XSF-TDA is provided and supports both the ALDA0 and multicollinear XC kernels. Extensions to energy gradients, nonadiabatic couplings, and spin-orbit couplings are underway and will be reported in due course.

\section*{Acknowledgment}
The authors acknowledge Yu Pang and Prof. Qian Peng for sharing the structures of INVEST-45 and helpful discussions. This work was supported by the Quantum Science and Technology-National Science and Technology Major Project (2023ZD0300200) and the Fundamental Research Funds for the Central Universities. 

\begin{suppinfo}
	
	The Supporting Information is available free of charge at
	
	\begin{quote}
		Detailed numerical data for Loos-10, QUEST-19, NAH159, MR-TADF-35, Heptazine-4, and INVEST-45 sets.
    \end{quote}
	
\end{suppinfo}

\section*{Data and Software Availability}
An open-source implementation of XSF-TDA is available in the \textsc{XTDDFT} package\cite{XTDDFTcode}. 
Data reported in this work can be found at Ref. \cite{xsftda-stgap}.

\FloatBarrier
\bibliography{reference}

\end{document}


\newpage
\tableofcontents
\newpage

\section{Loos-10 calibration set}

The vertical excitation energies (VEEs) obtained from USF-TDA and XSF-TDA for the Loos-10\cite{2023heptazine} 
calibration set are summarized in Tables \ref{tab:alda0_s0}-\ref{tab:loos-10-final}.

\begin{table}[!htbp]
    \centering\footnotesize
    \setlength{\tabcolsep}{3pt}
    \begin{tabular}{ccccccccccc}
    \hline\hline
        ~ & ~ & \multicolumn{4}{c}{USF-TDA(ALDA0)} && 
        \multicolumn{4}{c}{XSF-TDA(ALDA0)} \\ 
        \cmidrule(r){3-6} \cmidrule(r){8-11}
        No. & TBE\cite{2023heptazine}  & BLYP & B3LYP & PBE0 & BHHLYP & ~ & BLYP & B3LYP & PBE0 & BHHLYP \\ 
        \hline
        1 & 2.936  & 2.536  & 2.706  & 2.750  & 2.895  & ~ & 2.606  & 2.857  & 2.927  & 3.233  \\ 
        2 & 1.110  & 1.152  & 1.125  & 1.090  & 1.017  & ~ & 1.230  & 1.289  & 1.289  & 1.372  \\ 
        3 & 1.663  & 1.579  & 1.602  & 1.587  & 1.506  & ~ & 1.653  & 1.762  & 1.779  & 1.912  \\ 
        4 & 2.296  & 2.043  & 2.141  & 2.154  & 2.180  & ~ & 2.113  & 2.286  & 2.324  & 2.524  \\ 
        5 & 2.230  & 2.015  & 2.096  & 2.101  & 2.061  & ~ & 2.087  & 2.252  & 2.287  & 2.476  \\ 
        6 & 0.904  & 1.013  & 0.933  & 0.880  & 0.691  & ~ & 1.100  & 1.125  & 1.115  & 1.156  \\ 
        7 & 0.735  & 0.897  & 0.783  & 0.719  & 0.505  & ~ & 0.992  & 0.988  & 0.971  & 0.975  \\ 
        8 & 0.583  & 0.795  & 0.652  & 0.578  & 0.341  & ~ & 0.897  & 0.867  & 0.843  & 0.812  \\ 
        9 & 1.463  & 1.627  & 1.400  & 1.340  & 1.225  & ~ & 1.409  & 1.551  & 1.542  & 1.678  \\ 
        10 & 1.827  & 1.449  & 1.775  & 1.758  & 1.759  & ~ & 1.531  & 1.886  & 1.894  & 2.073  \\ 
        \hline\hline
    \end{tabular}
    \caption{VEEs (in eV) for S$_0\rightarrow$ T$_1$ computed using USF-TDA(ALDA0) and XSF-TDA(ALDA0) with the 6-31G* basis set.}\label{tab:alda0_s0}
\end{table}

\begin{table}[!htbp]
    \centering\footnotesize
    \begin{tabular}{cccccc}
    \hline\hline
        No.  & TBE\cite{2023heptazine}  & BLYP & B3LYP & PBE0 & BHHLYP \\ 
        \hline
        1 & 2.936  & 2.564  & 2.727  & 2.776  & 2.912  \\ 
        2 & 1.110  & 1.184  & 1.149  & 1.120  & 1.037  \\ 
        3 & 1.663  & 1.611  & 1.627  & 1.619  & 1.529  \\ 
        4 & 2.296  & 2.075  & 2.167  & 2.187  & 2.206  \\ 
        5 & 2.230  & 2.047  & 2.123  & 2.135  & 2.087  \\ 
        6 & 0.904  & 1.048  & 0.963  & 0.918  & 0.723  \\ 
        7 & 0.735  & 0.934  & 0.814  & 0.759  & 0.536  \\ 
        8 & 0.583  & 0.833  & 0.682  & 0.617  & 0.367  \\ 
        9 & 1.463  & 1.648  & 1.424  & 1.374  & 1.258  \\ 
        10 & 1.827  & 1.469  & 1.788  & 1.776  & 1.780  \\ 
        \hline\hline
    \end{tabular}
    \caption{VEEs (in eV) for S$_0\rightarrow$ T$_1$ computed using USF-TDA(MCOL) in the 6-31G* basis set.}
    \label{tab:usf_multicollinear_s0}
\end{table}

\begin{table}[!htbp]
    \centering\footnotesize
    \setlength{\tabcolsep}{3pt}
    \begin{tabular}{ccccccccccc}
    \hline\hline
    ~ & ~ & \multicolumn{4}{c}{$s_\mathrm{X}=1$} & ~ & 
    \multicolumn{4}{c}{$s_\mathrm{X}=4(c_{\mathrm{X}}-1/2)^2$} \\
        \cmidrule(r){3-6} \cmidrule(r){8-11}
        No.  & TBE\cite{2023heptazine}  & BLYP & B3LYP & PBE0 & BHHLYP & ~ & BLYP & B3LYP & PBE0 & BHHLYP \\ 
        \hline
        1 & 2.936  & 2.636  & 2.881  & 2.957  & 3.253  & ~ & 2.636  & 2.834  & 2.898  & 3.128  \\ 
        2 & 1.110  & 1.266  & 1.318  & 1.325  & 1.397  & ~ & 1.266  & 1.260  & 1.248  & 1.223  \\ 
        3 & 1.663  & 1.689  & 1.791  & 1.817  & 1.939  & ~ & 1.689  & 1.737  & 1.746  & 1.781  \\ 
        4 & 2.296  & 2.150  & 2.316  & 2.363  & 2.553  & ~ & 2.150  & 2.265  & 2.296  & 2.419  \\ 
        5 & 2.230  & 2.123  & 2.282  & 2.325  & 2.504  & ~ & 2.123  & 2.232  & 2.261  & 2.371  \\ 
        6 & 0.904  & 1.140  & 1.159  & 1.159  & 1.188  & ~ & 1.140  & 1.093  & 1.071  & 0.987  \\ 
        7 & 0.735  & 1.034  & 1.024  & 1.017  & 1.009  & ~ & 1.034  & 0.953  & 0.922  & 0.789  \\ 
        8 & 0.583  & 0.940  & 0.903  & 0.889  & 0.844  & ~ & 0.940  & 0.826  & 0.786  & 0.603  \\ 
        9 & 1.463  & 1.437  & 1.577  & 1.577  & 1.706  & ~ & 1.437  & 1.526  & 1.507  & 1.538  \\ 
        10 & 1.827  & 1.553  & 1.902  & 1.916  & 2.087  & ~ & 1.553  & 1.861  & 1.858  & 1.962  \\ 
        \hline\hline
    \end{tabular} 
    \caption{VEEs (in eV) for S$_0\rightarrow$ T$_1$ transitions computed using XSF-TDA(MCOL) in the 6-31G* basis set.
    }
\end{table}
\FloatBarrier

\begin{table}[!htbp]
    \centering\footnotesize
    \setlength{\tabcolsep}{3pt}
    \begin{tabular}{ccccccccccc}
    \hline\hline
        ~ & ~ & \multicolumn{4}{c}{USF-TDA(ALDA0)} & ~ &
        \multicolumn{4}{c}{XSF-TDA(ALDA0)} \\ 
        \cmidrule(r){3-6} \cmidrule(r){8-11}
        No.  & TBE\cite{2023heptazine} & BLYP & B3LYP & PBE0 & BHHLYP & ~ & BLYP & B3LYP & PBE0 & BHHLYP \\ 
        \hline
        1 & -0.219  & 0.139  & 0.075  & 0.069  & -0.151  & ~ & 0.019  & -0.275  & -0.335  & -0.845  \\ 
        2 & -0.131  & 0.134  & 0.087  & 0.083  & -0.059  & ~ & 0.045  & -0.205  & -0.249  & -0.678  \\ 
        3 & -0.101  & 0.139  & 0.108  & 0.115  & 0.043  & ~ & 0.055  & -0.179  & -0.217  & -0.598  \\ 
        4 & -0.119  & 0.140  & 0.109  & 0.122  & 0.074  & ~ & 0.056  & -0.176  & -0.209  & -0.525  \\ 
        5 & -0.103  & 0.143  & 0.115  & 0.128  & 0.082  & ~ & 0.054  & -0.176  & -0.208  & -0.532  \\ 
        6 & -0.071  & 0.142  & 0.108  & 0.111  & 0.047  & ~ & 0.068  & -0.162  & -0.198  & -0.585  \\ 
        7 & -0.042  & 0.148  & 0.116  & 0.120  & 0.055  & ~ & 0.081  & -0.147  & -0.182  & -0.573  \\ 
        8 & -0.029  & 0.152  & 0.118  & 0.119  & 0.017  & ~ & 0.090  & -0.143  & -0.179  & -0.585  \\ 
        9 & -0.199  & -0.358  & 0.076  & 0.111  & -0.018  & ~ & -0.057  & -0.227  & -0.252  & -0.607  \\ 
        10 & -0.305  & -0.104  & 0.025  & 0.038  & -0.182  & ~ & -0.168  & -0.332  & -0.382  & -0.884  \\ 
        \hline\hline
    \end{tabular}
    \caption{$\Delta E_{\mathrm{ST}}$ (in eV) computed using USF-TDA and XSF-TDA with the ALDA0 kernel in the 6-31G* basis set.}
\end{table}

\begin{table}[!htbp]
    \centering\footnotesize
    \begin{tabular}{cccccc}
    \hline\hline
        No.  & TBE\cite{2023heptazine} & BLYP & B3LYP & PBE0 & BHHLYP \\ 
        \hline
        1 & -0.219 & 0.018  & -0.027  & -0.053  & -0.239  \\ 
        2 & -0.131 & 0.018  & -0.005  & -0.024  & -0.142  \\ 
        3 & -0.101 & 0.020  & 0.015  & 0.006  & -0.038  \\ 
        4 & -0.119 & 0.019  & 0.016  & 0.011  & -0.003  \\ 
        5 & -0.103 & 0.020  & 0.019  & 0.014  & 0.000  \\ 
        6 & -0.071 & 0.022  & 0.019  & 0.009  & -0.030  \\ 
        7 & -0.042 & 0.024  & 0.025  & 0.017  & -0.023  \\ 
        8 & -0.029 & 0.026  & 0.025  & 0.016  & -0.063  \\ 
        9 & -0.199 & -0.397  & -0.017  & -0.027  & -0.091  \\ 
        10 & -0.305 & -0.179  & -0.085  & -0.118  & -0.267  \\ 
        \hline\hline
    \end{tabular}
    \caption{$\Delta E_{\mathrm{ST}}$ (in eV) computed using USF-TDA(MCOL) in the 6-31G* basis set.}
\end{table}

\begin{table}[!htbp]
    \centering\footnotesize
    \setlength{\tabcolsep}{3pt}
    \begin{tabular}{ccccccccccc}
    \hline\hline
    ~ & ~ & \multicolumn{4}{c}{$s_\mathrm{X}=1$} & ~ & 
    \multicolumn{4}{c}{$s_\mathrm{X}=4(c_{\mathrm{X}}-1/2)^2$} \\
        \cmidrule(r){3-6} \cmidrule(r){8-11}
        No.  & TBE\cite{2023heptazine}  & BLYP & B3LYP & PBE0 & BHHLYP & ~ & BLYP & B3LYP & PBE0 & BHHLYP \\ 
        \hline
        1 & -0.219  & -0.161  & -0.429  & -0.511  & -0.978  & ~ & -0.161  & -0.171  & -0.174  & -0.274  \\ 
        2 & -0.131  & -0.129  & -0.353  & -0.420  & -0.803  & ~ & -0.129  & -0.132  & -0.131  & -0.206  \\ 
        3 & -0.101  & -0.116  & -0.324  & -0.384  & -0.720  & ~ & -0.116  & -0.097  & -0.086  & -0.097  \\ 
        4 & -0.119  & -0.114  & -0.319  & -0.374  & -0.644  & ~ & -0.114  & -0.087  & -0.070  & -0.006  \\ 
        5 & -0.103  & -0.119  & -0.323  & -0.378  & -0.654  & ~ & -0.119  & -0.086  & -0.067  & 0.003  \\ 
        6 & -0.071  & -0.100  & -0.305  & -0.363  & -0.706  & ~ & -0.100  & -0.085  & -0.074  & -0.107  \\ 
        7 & -0.042  & -0.084  & -0.288  & -0.344  & -0.693  & ~ & -0.084  & -0.069  & -0.056  & -0.094  \\ 
        8 & -0.029  & -0.076  & -0.284  & -0.341  & -0.707  & ~ & -0.076  & -0.067  & -0.057  & -0.120  \\ 
        9 & -0.199  & -0.200  & -0.377  & -0.445  & -0.728  & ~ & -0.200  & -0.142  & -0.135  & -0.102  \\ 
        10 & -0.305  & -0.245  & -0.489  & -0.584  & -1.013  & ~ & -0.245  & -0.238  & -0.253  & -0.343  \\ 
        \hline\hline
    \end{tabular} 
    \caption{$\Delta E_{\mathrm{ST}}$ (in eV) computed using XSF-TDA(MCOL) in the 6-31G* basis set.}
    \label{tab:loos-10-final}
\end{table}

\clearpage
\section{QUEST-19 validation set}

\begin{figure*}[!tp]
\centering
\includegraphics[width=0.7\linewidth]{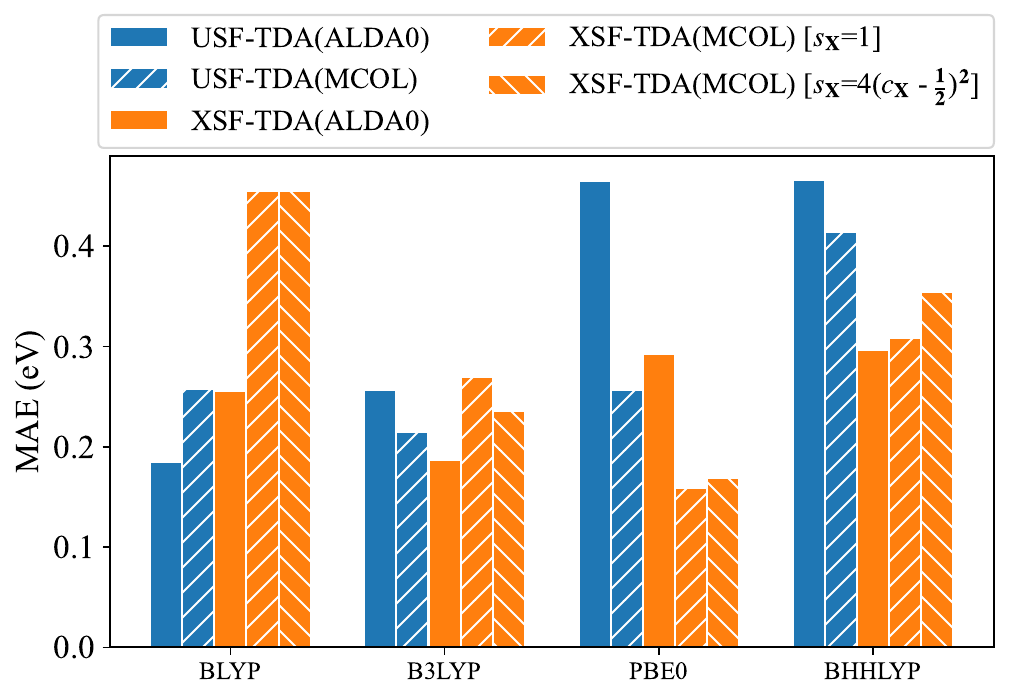}
\caption{MAEs of VEEs (in eV) for D$_0\rightarrow$ Q$_1$ obtained from USF-TDA and XSF-TDA.}
\label{fig:quest-19}
\end{figure*}

The vertical excitation energies (VEEs) for D$_0\rightarrow$ Q$_1$ in QUEST-19, computed using USF-TDA and XSF-TDA with a Q$_1$ reference and the aug-cc-pVTZ basis set, are summarized in Tables \ref{tab:quest-19a}--\ref{tab:quest-19c}.
The reported energies are $\Delta E_{\mathrm{DQ}}=E(\mathrm{Q}_1)-E(\mathrm{D}_0)=-\omega_0$, where $\omega_0$ is the negative Q$_1\rightarrow$ D$_0$ response root from the quartet reference. The corresponding MAEs are displayed in Fig. \ref{fig:quest-19}.


\begin{table}[!ht]
    \centering
    \setlength{\tabcolsep}{3pt}
    \begin{tabular}{ccccccccccc}
    \hline\hline
     ~ & ~ & \multicolumn{4}{c}{USF-TDA(ALDA0)} && 
        \multicolumn{4}{c}{XSF-TDA(ALDA0)} \\ 
     \cmidrule(r){3-6} \cmidrule(r){8-11}
        Molecule & TBE\cite{Loos2025The} & BLYP & B3LYP & PBE0 & BHHLYP & ~ & BLYP & B3LYP & PBE0 & BHHLYP \\ 
        \hline
        \ce{AlCH2} & 2.096  & 2.129  & 2.156  & 1.835  & 2.233  & ~ & 2.163  & 2.210  & 1.878  & 2.326  \\ 
        \ce{Allyl} & 6.012  & 5.100  & 5.203  & 4.986  & 5.281  & ~ & 5.575  & 7.102  & 7.297  & 5.532  \\ 
        \ce{BH2} & 5.373  & 5.425  & 5.530  & 5.230  & 5.689  & ~ & 5.483  & 5.619  & 5.319  & 5.844  \\ 
        \ce{CH2N} & 5.296  & 5.118  & 5.058  & 4.785  & 4.917  & ~ & 5.270  & 5.248  & 5.023  & 5.187  \\ 
        \ce{CH2O+} & 6.928  & 6.821  & 6.555  & 6.233  & 6.162  & ~ & 6.988  & 6.768  & 6.494  & 6.478  \\ 
        \ce{ClO2} & 6.239  & 6.046  & 5.929  & 5.790  & 5.336  & ~ & 6.190  & 6.218  & 6.126  & 5.893  \\ 
        \ce{Ethynyl} & 5.184  & 5.276  & 5.260  & 4.994  & 5.175  & ~ & 6.327  & 5.423  & 5.199  & 5.396  \\ 
        \ce{F2BS} & 6.400  & 6.470  & 6.446  & 6.118  & 6.254  & ~ & 6.950  & 6.579  & 6.295  & 6.463  \\ 
        \ce{HBCl} & 5.945  & 5.879  & 5.924  & 5.734  & 5.835  & ~ & 5.953  & 6.058  & 5.894  & 6.109  \\ 
        \ce{H2BO} & 6.923  & 6.843  & 6.395  & 5.956  & 5.794  & ~ & 6.997  & 6.585  & 6.183  & 6.055  \\ 
        \ce{HCO} & 6.389  & 6.377  & 6.176  & 5.880  & 5.830  & ~ & 6.522  & 6.416  & 6.164  & 6.250  \\ 
        \ce{HOC} & 3.856  & 3.650  & 3.666  & 3.168  & 3.905  & ~ & 3.724  & 3.779  & 3.278  & 4.071  \\ 
        \ce{H2PO} & 6.325  & 6.312  & 6.103  & 5.782  & 5.480  & ~ & 6.431  & 6.320  & 6.039  & 5.886  \\ 
        \ce{H2PS} & 5.138  & 4.976  & 4.924  & 4.665  & 4.764  & ~ & 5.061  & 5.061  & 4.848  & 4.998  \\ 
        \ce{NH2} & 7.340  & 8.226  & 7.862  & 7.298  & 7.317  & ~ & 8.343  & 8.036  & 7.496  & 7.601  \\ 
        \ce{Nitromethyl} & 4.340  & 4.334  & 4.109  & 3.931  & 3.331  & ~ & 4.486  & 4.401  & 4.268  & 3.854  \\ 
        \ce{NO2} & 4.617  & 4.678  & 4.274  & 4.146  & 3.561  & ~ & 4.866  & 4.635  & 4.566  & 4.292  \\ 
        \ce{PH2} & 6.148  & 5.950  & 6.039  & 5.697  & 6.185  & ~ & 6.595  & 6.142  & 5.815  & 6.352  \\ 
        \ce{Vinyl} & 4.561  & 4.376  & 4.352  & 4.047  & 4.280  & ~ & 4.520  & 4.529  & 4.276  & 4.525  \\ 
        \hline\hline
    \end{tabular}
    \caption{VEEs (in eV) for D$_0\rightarrow$ Q$_1$ computed using USF-TDA(ALDA0) and XSF-TDA(ALDA0) with the aug-cc-pVTZ basis set starting from Q$_1$ as reference.}\label{tab:quest-19a}
\end{table}

\begin{table}[!ht]
    \centering
    \begin{tabular}{ccccccc}
    \hline\hline
         Molecule& TBE\cite{Loos2025The} & ~ & BLYP & B3LYP & PBE0 & BHHLYP \\ 
         \hline
        \ce{AlCH2} & 2.096  & ~ & 2.358  & 2.341  & 2.106  & 2.406  \\ 
        \ce{Allyl} & 6.012  & ~ & 5.334  & 5.436  & 5.305  & 5.574  \\ 
        \ce{BH2} & 5.373  & ~ & 5.672  & 5.713  & 5.453  & 5.829  \\ 
        \ce{CH2N} & 5.296  & ~ & 5.545  & 5.392  & 5.182  & 5.195  \\ 
        \ce{CH2O+} & 6.928  & ~ & 7.224  & 6.881  & 6.611  & 6.444  \\ 
        \ce{ClO2} & 6.239  & ~ & 6.223  & 6.035  & 5.909  & 5.429  \\ 
        \ce{Ethynyl} & 5.184  & ~ & 5.705  & 5.593  & 5.392  & 5.455  \\ 
        \ce{F2BS} & 6.400  & ~ & 6.614  & 6.567  & 6.281  & 6.367  \\ 
        \ce{HBCl} & 5.945  & ~ & 6.000  & 6.012  & 5.841  & 5.899  \\ 
        \ce{H2BO} & 6.923  & ~ & 7.076  & 6.591  & 6.204  & 5.968  \\ 
        \ce{HCO} & 6.389  & ~ & 6.594  & 6.351  & 6.080  & 5.984  \\ 
        \ce{HOC} & 3.856  & ~ & 4.135  & 4.057  & 3.660  & 4.252  \\ 
        \ce{H2PO} & 6.325  & ~ & 6.398  & 6.167  & 5.866  & 5.529  \\ 
        \ce{H2PS} & 5.138  & ~ & 5.082  & 5.010  & 4.771  & 4.836  \\ 
        \ce{NH2} & 7.340  & ~ & 8.271  & 7.892  & 7.362  & 7.335  \\ 
        \ce{Nitromethyl} & 4.340  & ~ & 4.425  & 4.285  & 4.137  & 3.519  \\ 
        \ce{NO2} & 4.617  & ~ & 4.813  & 4.387  & 4.267  & 3.665  \\ 
        \ce{PH2} & 6.148  & ~ & 6.252  & 6.279  & 5.999  & 6.406  \\ 
        \ce{Vinyl} & 4.561  & ~ & 4.779  & 4.666  & 4.441  & 4.537  \\ 
        \hline\hline
    \end{tabular}
    \caption{VEEs (in eV) for D$_0\rightarrow$ Q$_1$ computed using USF-TDA(MCOL) with the aug-cc-pVTZ basis set starting from Q$_1$ as reference.}\label{tab:quest-19b}
\end{table}

\begin{table}[!ht]
    \centering
    \setlength{\tabcolsep}{3pt}
    \begin{tabular}{ccccccccccc}
    \hline\hline
        ~ & ~ & \multicolumn{4}{c}{$s_\mathrm{X}=1$} & ~ & 
        \multicolumn{4}{c}{$s_\mathrm{X}=4(c_{\mathrm{X}}-1/2)^2$} \\
        \cmidrule(r){3-6} \cmidrule(r){8-11}
        Molecule & TBE\cite{Loos2025The} & BLYP & B3LYP & PBE0 & BHHLYP & ~ & BLYP & B3LYP & PBE0 & BHHLYP \\
        \hline
        \ce{AlCH2} & 2.096  & 2.395  & 2.398  & 2.158  & 2.502  & ~ & 2.395  & 2.381  & 2.137  & 2.462  \\ 
        \ce{Allyl} & 6.012  & 5.441  & 5.590  & 5.484  & 5.822  & ~ & 5.441  & 5.559  & 5.443  & 5.738  \\ 
        \ce{BH2} & 5.373  & 5.734  & 5.806  &  5.548  & 5.989  & ~ & 5.734  & 5.784  & 5.521  & 5.931  \\ 
        \ce{CH2N} & 5.296  & 5.694  & 5.580  &  5.419  & 5.464  & ~ & 5.694  & 5.545  & 5.373  & 5.369  \\ 
        \ce{CH2O+} & 6.928  & 7.389  & 7.091  &  6.869  & 6.758  & ~ & 7.389  & 7.049  & 6.815  & 6.639  \\ 
        \ce{ClO2} & 6.239  & 6.370  & 6.330  &  6.251  & 5.991  & ~ & 6.370  & 6.241  & 6.131  & 5.680  \\ 
        \ce{Ethynyl} & 5.184  & 6.752  & 5.754  &  5.595  & 5.676  & ~ & 6.752  & 5.718  & 5.548  & 5.581  \\ 
        \ce{F2BS} & 6.400  & 7.097  & 6.702  & 6.461  & 6.578  & ~ & 7.097  & 6.677  & 6.428  & 6.501  \\ 
        \ce{HBCl} & 5.945  & 6.076  & 6.149  & 6.004  & 6.175  & ~ & 6.076  & 6.116  & 5.958  & 6.042  \\ 
        \ce{H2BO} & 6.923  & 7.229  & 6.780  & 6.430  & 6.229  & ~ & 7.229  & 6.729  & 6.365  & 6.081  \\ 
        \ce{HCO} & 6.389  & 6.742  & 6.593  &  6.367  & 6.405  & ~ & 6.742  & 6.508  & 6.250  & 6.139  \\ 
        \ce{HOC} & 3.856  & 4.210  & 4.170  &  3.777  & 4.419  & ~ & 4.210  & 4.131  & 3.725  & 4.326  \\ 
        \ce{H2PO} & 6.325  & 6.520  & 6.388  & 6.127  & 5.937  & ~ & 6.520  & 6.309  & 6.016  & 5.640  \\ 
        \ce{H2PS} & 5.138  & 5.171  & 5.150  &  4.959  & 5.073  & ~ & 5.171  & 5.106  & 4.897  & 4.931  \\ 
        \ce{NH2} & 7.340  & 8.388  & 8.066  &  7.563  & 7.617  & ~ & 8.388  & 7.988  & 7.460  & 7.390  \\ 
        \ce{Nitromethyl} & 4.340  & 4.580  & 4.579  & 4.476  & 4.042  & ~ & 4.580  & 4.528  & 4.409  & 3.882  \\ 
        \ce{NO2} & 4.617  & 5.007  & 4.754  & 4.693  & 4.401  & ~ & 5.007  & 4.657  & 4.561  & 4.055  \\ 
        \ce{PH2} & 6.148  & 6.900  & 6.385  & 6.122  & 6.576  & ~ & 6.900  & 6.357  & 6.084  & 6.503  \\ 
        \ce{Vinyl} & 4.561  & 4.920  & 4.841  &  4.668  & 4.781  & ~ & 4.920  & 4.807  & 4.624  & 4.692  \\ 
        \hline\hline
    \end{tabular} 
    \caption{VEEs (in eV) for D$_0\rightarrow$ Q$_1$ computed using XSF-TDA(MCOL) with the aug-cc-pVTZ basis set starting from Q$_1$ as reference.}\label{tab:quest-19c}
\end{table}

\FloatBarrier

\clearpage

\section{NAH159 benchmark set}
Calculated $\Delta E_{\mathrm{ST}}$ using the B3LYP and PBE0 functionals are provided in Tables \ref{tab:nah159_b3lyp} and \ref{tab:nah159_pbe0}, respectively. The CV(1) weights obtained for the S$_1$ state
in the XSF-TDA calculations with these functionals are summarized in Table \ref{tab:nah159_cv1}.

{\singlespacing\footnotesize
\setlength{\tabcolsep}{3pt}
\begin{longtable}{ccc p{1.2cm}p{1.7cm}p{1.2cm}p{1.2cm}p{1.2cm}p{1.2cm}}
\caption{Calculated $\Delta E_{\mathrm{ST}}$ (in eV) with B3LYP/def2-SVP for the NAH159 benchmark set.} \\
\hline
\hline

        ~ & ~ & ~ & ~ & ~ & ALDA0 & ~ & MCOL &   \\ \cmidrule(r){6-7} \cmidrule(r){8-9}
        Molecule & Ref.\cite{kunze2024} & TDA & TD-DFT & USF-TD-DFT & USF-TDA & XSF-TDA & USF-TDA & XSF-TDA  \\ 
        \hline
\endfirsthead
\multicolumn{9}{c}{\tablename~\thetable{} (continued)} \\
\hline
\hline

        ~ & ~ & ~ & ~ & ~ & ALDA0 & ~ & MCOL &   \\ \cmidrule(r){6-7} \cmidrule(r){8-9}
        Molecule & Ref.\cite{kunze2024} & TDA & TD-DFT & USF-TD-DFT & USF-TDA & XSF-TDA & USF-TDA & XSF-TDA  \\ 
        \hline
\endhead
\hline
\multicolumn{9}{r}{Continued on next page} \\
\endfoot
\endlastfoot
        azulene\_2nh2 & 0.358  & 0.783  & 0.967  & 0.753  & 0.432  & 0.431  & 0.357  & 0.365   \\ 
        azulene\_2nh2\_4me & 0.112  & 0.525  & 0.725  & 0.833  & 0.169  & 0.188  & 0.106  & 0.121   \\ 
        azulene\_2oh\_2me & -0.016  & 0.419  & 0.384  & 0.012  & 0.345  & 0.012  & 0.221  & 0.055   \\ 
        azulene\_2oh\_2nh2 & 0.309  & 0.758  & 0.915  & 0.580  & 0.355  & 0.349  & 0.270  & 0.265   \\ 
        azulene\_2oh\_alt & -0.034  & 0.408  & 0.375  & 0.018  & 0.372  & 0.010  & 0.239  & 0.057   \\ 
        azulene\_2ome\_2nh2 & 0.370  & 0.795  & 0.934  & 0.697  & 0.445  & 0.428  & 0.365  & 0.363   \\ 
        azulene\_2ome\_alt & -0.032  & 0.392  & 0.358  & 0.017  & 0.362  & 0.009  & 0.225  & 0.044   \\ 
        azulene\_4cf3 & -0.064  & 0.401  & 0.388  & 0.017  & 0.367  & 0.015  & 0.226  & 0.041   \\
\end{longtable} 

\begin{longtable}{ccc p{1.2cm}p{1.6cm}p{1.2cm}p{1.2cm}p{1.2cm}p{1.2cm}}
 \caption{Calculated $\Delta E_{\mathrm{ST}}$ (in eV) with PBE0/def2-SVP for the NAH159 benchmark set.} \\
\hline
\hline
        ~ & ~ & ~ & ~ & ~ & ALDA0 & ~ & MCOL &   \\ \cmidrule(r){6-7} \cmidrule(r){8-9}
        Molecule & Ref.\cite{kunze2024} & TDA & TD-DFT & USF-TD-DFT & USF-TDA & XSF-TDA & USF-TDA & XSF-TDA  \\ 
        \hline
\endfirsthead
\multicolumn{9}{c}{\tablename~\thetable{} (continued)} \\
\hline
\hline
        ~ & ~ & ~ & ~ & ~ & ALDA0 & ~ & MCOL &   \\ \cmidrule(r){6-7} \cmidrule(r){8-9}
        Molecule & Ref.\cite{kunze2024} & TDA & TD-DFT & USF-TD-DFT & USF-TDA & XSF-TDA & USF-TDA & XSF-TDA  \\ 
        \hline
\endhead
\hline
\multicolumn{9}{r}{Continued on next page} \\
\endfoot
\endlastfoot
        azulene\_2nh2 & 0.358  & 0.849  & 1.133  & 0.821  & 0.455  & 0.455  & 0.361  & 0.379   \\ 
        azulene\_2nh2\_4me & 0.112  & 0.589  & 0.896  & 0.891  & 0.196  & 0.213  & 0.117  & 0.139   \\ 
        azulene\_2oh\_2me & -0.016  & 0.447  & 0.420  & -0.032  & 0.401  & 0.012  & 0.249  & 0.083   \\ 
        azulene\_2oh\_2nh2 & 0.309  & 0.811  & 1.047  & 0.639  & 0.376  & 0.365  & 0.274  & 0.269   \\ 
        azulene\_2oh\_alt & -0.034  & 0.438  & 0.525  & -0.028  & 0.440  & 0.012  & 0.271  & 0.085   \\ 
        azulene\_2ome\_2nh2 & 0.370  & 0.850  & 1.059  & 0.755  & 0.463  & 0.445  & 0.366  & 0.372   \\ 
        azulene\_2ome\_alt & -0.032  & 0.421  & 0.430  & -0.023  & 0.430  & 0.012  & 0.256  & 0.067   \\ 
        azulene\_4cf3 & -0.064  & 0.426  & 0.616  & -0.024  & 0.436  & 0.017  & 0.258  & 0.063   \\
\end{longtable}

\begin{longtable}{ccccc}
\caption{CV(1) percentage (\%) for the S$_1$ state in XSF-TDA calculations for the NAH159 benchmark set.}\\

\hline
\hline
        ~ & B3LYP & ~ & PBE0 & ~ \\ \cmidrule(r){2-3} \cmidrule(r){4-5}
        Molecule & ALDA0 & MCOL & ALDA0 & MCOL \\ 
        \hline
\endfirsthead
\multicolumn{5}{c}{\tablename~\thetable{} (continued)} \\
\hline
\hline
        ~ & B3LYP & ~ & PBE0 & ~ \\ \cmidrule(r){2-3} \cmidrule(r){4-5}
        Molecule & ALDA0 & MCOL & ALDA0 & MCOL \\ 
        \hline
\endhead
\hline
\multicolumn{5}{r}{Continued on next page} \\
\endfoot
\endlastfoot
        azulene\_2nh2 & 2.839  & 2.985  & 3.488  & 3.706  \\ 
        azulene\_2nh2\_4me & 2.675  & 2.711  & 3.297  & 3.350  \\ 
        azulene\_2oh\_2me & 3.001  & 2.161  & 3.416  & 2.317  \\ 
        azulene\_2oh\_2nh2 & 2.850  & 2.920  & 3.496  & 3.592  \\ 
        azulene\_2oh\_alt & 3.098  & 2.246  & 3.520  & 2.406  \\ 
        azulene\_2ome\_2nh2 & 2.767  & 2.898  & 3.393  & 3.576  \\ 
        azulene\_2ome\_alt & 2.912  & 2.089  & 3.321  & 2.238  \\ 
        azulene\_4cf3 & 2.913  & 2.091  & 3.346  & 2.267  \\
\end{longtable}
}

\clearpage
\section{MR-TADF-35 benchmark set}

Calculated $\Delta E_{\mathrm{ST}}$ with the B3LYP and PBE0 functionals are provided in Tables \ref{tab:mewes_b3lyp} and \ref{tab:mewes_pbe0}, respectively. In the XSF-TDA calculation, the CV(1) percentages for the S$_1$ state obtained with these functionals summarized in Table \ref{tab:mewes_cv1}.

\begin{table}[!htbp]
    \centering\footnotesize
    \setlength{\tabcolsep}{4pt}
    \begin{tabular}{ccc p{1.2cm}p{1.7cm}p{1.2cm}p{1.2cm}p{1.2cm}p{1.2cm}}
    \hline\hline
         ~ & ~ & ~ & ~ & ~ & ALDA0 & ~ & MCOL&   \\ \cmidrule(r){6-7} \cmidrule(r){8-9}
        Molecule & Exp.\cite{kunze2025} & TDA & TD-DFT & USF-TD-DFT & USF-TDA & XSF-TDA & USF-TDA & XSF-TDA  \\ 
        \hline
        DOBNA & 0.180  & 0.579  & 0.521  & 0.196  & 0.392  & 0.135  & 0.317  & 0.177   \\ 
        DABNA-1 & 0.150  & 0.540  & 0.486  & 0.169  & 0.375  & 0.099  & 0.301  & 0.150   \\ 
        DABNA-2 & 0.150  & 0.434  & 0.402  & 0.125  & 0.301  & 0.063  & 0.243  & 0.112   \\ 
        t-DABNA & 0.160  & 0.534  & 0.477  & 0.166  & 0.366  & 0.097  & 0.294  & 0.146   \\ 
        BN-DMAC & 0.140  & 0.461  & 0.415  & 0.126  & 0.313  & 0.063  & 0.244  & 0.104   \\ 
        TABNA & 0.210  & 0.540  & 0.515  & 0.221  & 0.182  & 0.141  & 0.157  & 0.166   \\
\hline\hline
    \end{tabular}
    \caption{Calculated $\Delta E_{\mathrm{ST}}$ (in eV) with B3LYP/def2-SVP for the MR-TADF-35 benchmark.}
    \label{tab:mewes_b3lyp}
\end{table}

\begin{table}[!htbp]
    \centering\footnotesize
    \setlength{\tabcolsep}{4pt}
    \begin{tabular}{ccc p{1.2cm}p{1.7cm}p{1.2cm}p{1.2cm}p{1.2cm}p{1.2cm}}
    \hline\hline
         ~ & ~ & ~ & ~ & ~ & ALDA0 & ~ & MCOL&   \\ \cmidrule(r){6-7} \cmidrule(r){8-9}
        Molecule & Exp. & TDA & TD-DFT & USF-TD-DFT & USF-TDA & XSF-TDA & USF-TDA & XSF-TDA  \\ 
        \hline
        DOBNA & 0.180  & 0.624  & 0.573  & 0.198  & 0.475  & 0.161  & 0.370  & 0.216   \\ 
        DABNA-1 & 0.150  & 0.582  & 0.533  & 0.165  & 0.451  & 0.118  & 0.351  & 0.187   \\ 
        DABNA-2 & 0.150  & 0.480  & 0.456  & 0.122  & 0.374  & 0.078  & 0.294  & 0.146   \\ 
        t-DABNA & 0.160  & 0.575  & 0.524  & 0.163  & 0.440  & 0.115  & 0.343  & 0.181   \\ 
        BN-DMAC & 0.140  & 0.494  & 0.457  & 0.122  & 0.384  & 0.079  & 0.289  & 0.134   \\ 
        TABNA & 0.210  & 0.594  & 0.589  & 0.228  & 0.267  & 0.169  & 0.237  & 0.228   \\
\hline\hline
    \end{tabular}
    \caption{Calculated $\Delta E_{\mathrm{ST}}$ (in eV) with PBE0/def2-SVP for the MR-TADF-35 benchmark.}
    \label{tab:mewes_pbe0}
\end{table}

\begin{table}[!htbp]
    \centering\footnotesize
    \begin{tabular}{ccccc}
    \hline\hline
        ~ & B3LYP & ~ & PBE0 & ~ \\ \cmidrule(r){2-3} \cmidrule(r){4-5}
        Molecule & ALDA0 & MCOL& ALDA0 & MCOL\\ 
        \hline
        DOBNA & 2.193  & 1.518  & 2.610  & 1.708  \\ 
        DABNA-1 & 2.377  & 1.648  & 2.818  & 1.848  \\ 
        DABNA-2 & 2.324  & 1.565  & 2.783  & 1.778  \\ 
        t-DABNA & 2.319  & 1.600  & 2.755  & 1.796  \\ 
        BN-DMAC & 2.155  & 1.460  & 2.557  & 1.629  \\ 
        TABNA & 2.478  & 1.301  & 2.928  & 1.628  \\
\hline\hline
    \end{tabular}
    \caption{CV(1) percentage (\%) for the S$_1$ state in XSF-TDA calculations for the MR-TADF-35 benchmark.}
    \label{tab:mewes_cv1}
\end{table}

\clearpage
\section{Heptazine-4 benchmark set}

In the XSF-TDA calculations, the CV(1) percentages for the S$_1$ state obtained with B3LYP and PBE0 are summarized in Table \ref{tab:mija_cv1}.

\begin{table}[!htbp]
    \centering\footnotesize
    \begin{tabular}{ccccc}
    \hline
    \hline
        ~ & B3LYP & ~ & PBE0 & ~ \\ \cmidrule(r){2-3} \cmidrule(r){4-5}
        ~ & ALDA0 & MCOL& ALDA0 & MCOL\\ 
        \hline
        \ce{HzTFEX2} & 4.097 & 2.854 & 4.713 & 3.067 \\ 
        \ce{HzPipX2} & 3.924 & 2.730 & 4.484 & 2.930 \\ 
        \ce{HzTFEP2} & 4.139 & 2.888 & 4.745 & 3.107 \\ 
        \ce{HzTFET2} & 4.049 & 2.831 & 4.661 & 3.054 \\ 
        \hline
        \hline
    \end{tabular}
    \caption{CV(1) percentage (\%) for the S$_1$ state in XSF-TDA calculations for the Heptazine-4 benchmark.}
    \label{tab:mija_cv1}
\end{table}

\clearpage
\section{INVEST-45 benchmark set}
Pang et al. reported 41 INVEST candidates from their descriptor-assisted screening method (DASM) and 46 INVEST molecules from their full-calculation screening method (FCSM).\cite{peng2025} The present calculations use a Pang-derived 45-molecule INVEST subset, denoted INVEST-45. The molecule numbers 1--45 in the tables below are internal labels for this subset and are not, by themselves, the M1--M45 identifiers in the Pang paper. The $\Delta E_{\mathrm{ST}}$ computed with the B3LYP and PBE0 functionals are provided in Tables \ref{tab:peng_b3lyp} and \ref{tab:peng_pbe0}, respectively. In the XSF-TDA calculation, the CV(1) percentages for the S$_1$ state obtained with these functionals are summarized in Table \ref{tab:peng_cv1}. 
Table \ref{tab:peng_ref} compiles the $\Delta E_{\mathrm{ST}}$ from SC-NEVPT2(8,8)/cc-pVDZ, CC2/aug-cc-pVDZ, and ADC(2)/aug-cc-pVDZ. The SC-NEVPT2(8,8)/cc-pVDZ results are taken from the work of Pang et al.,\cite{peng2025} while the CC2 and ADC(2) calculations were performed in this study. As shown in Table \ref{tab:peng_ref}, the CC2 and ADC(2) results exhibit a high level of consistency. Compared to the CC2 values, the mean absolute error (MAE) of the ADC(2) results is 0.008 eV, calculated using the unrounded energy gaps.

{\singlespacing\footnotesize
\setlength{\tabcolsep}{3pt}
\begin{longtable}{ccc p{1.2cm}p{1.7cm}p{1.2cm}p{1.2cm}p{1.2cm}p{1.2cm}}
\caption{Calculated $\Delta E_{\mathrm{ST}}$ (in eV) with B3LYP/aug-cc-pVDZ for the INVEST-45 subset.} \\
\hline
\hline

        ~ & ~ & ~ & ~ & ~ & ALDA0 & ~ & MCOL&   \\ \cmidrule(r){6-7} \cmidrule(r){8-9}
        Molecule & CC2 & TDA & TD-DFT & USF-TD-DFT & USF-TDA & XSF-TDA & USF-TDA & XSF-TDA  \\ 
        \hline
\endfirsthead
\multicolumn{9}{c}{\tablename~\thetable{} (continued)} \\
\hline
\hline

        ~ & ~ & ~ & ~ & ~ & ALDA0 & ~ & MCOL&   \\ \cmidrule(r){6-7} \cmidrule(r){8-9}
        Molecule & CC2 & TDA & TD-DFT & USF-TD-DFT & USF-TDA & XSF-TDA & USF-TDA & XSF-TDA  \\ 
        \hline
\endhead
\hline
\multicolumn{9}{r}{Continued on next page} \\
\endfoot
\endlastfoot

        1 & -0.215  & 0.199  & 0.214  & -0.276  & 0.224  & -0.221  & -0.007  & -0.135   \\ 
        2 & -0.100  & 0.240  & 0.253  & -0.188  & 0.212  & -0.155  & 0.106  & -0.080   \\ 
        3 & -0.117  & 0.241  & 0.251  & -0.242  & 0.218  & -0.175  & 0.118  & -0.086   \\ 
        4 & -0.192  & 0.199  & 0.205  & -0.285  & 0.210  & -0.219  & -0.010  & -0.135   \\ 
        5 & -0.145  & 0.198  & 0.199  & -0.286  & 0.146  & -0.201  & 0.056  & -0.116   \\ 
        6 & -0.093  & 0.232  & 0.242  & -0.199  & 0.207  & -0.158  & 0.117  & -0.070   \\ 
        7 & -0.076  & 0.248  & 0.268  & -0.168  & 0.264  & -0.164  & 0.149  & -0.091   \\ 
        8 & -0.135  & 0.212  & 0.221  & -0.247  & 0.193  & -0.185  & 0.099  & -0.103   \\ 
        9 & -0.138  & 0.222  & 0.232  & -0.213  & 0.215  & -0.179  & 0.107  & -0.104   \\ 
        10 & -0.093  & 0.214  & 0.223  & -0.238  & 0.191  & -0.178  & 0.096  & -0.105   \\ 
        11 & -0.137  & 0.209  & 0.219  & -0.216  & 0.201  & -0.178  & 0.099  & -0.109   \\ 
        12 & -0.093  & 0.203  & 0.214  & -0.190  & 0.216  & -0.161  & 0.104  & -0.095   \\ 
        13 & -0.090  & 0.215  & 0.233  & -0.180  & 0.232  & -0.159  & 0.121  & -0.098   \\ 
        14 & -0.095  & 0.192  & 0.205  & -0.213  & 0.200  & -0.172  & 0.097  & -0.102   \\ 
        15 & -0.089  & 0.207  & 0.225  & -0.182  & 0.209  & -0.163  & 0.110  & -0.090   \\ 
        16 & -0.321  & 0.103  & 0.107  & -0.478  & 0.030  & -0.330  & -0.077  & -0.235   \\ 
        17 & -0.156  & 0.199  & 0.205  & -0.262  & 0.204  & -0.197  & 0.098  & -0.118   \\ 
        18 & -0.133  & 0.215  & 0.225  & -0.246  & 0.196  & -0.183  & 0.097  & -0.104   \\ 
        19 & -0.152  & 0.193  & 0.199  & -0.270  & 0.174  & -0.201  & 0.002  & -0.125   \\ 
        20 & -0.069  & 0.198  & 0.216  & -0.163  & 0.218  & -0.142  & 0.108  & -0.083   \\ 
        21 & -0.080  & 0.213  & 0.239  & -0.149  & 0.237  & -0.154  & 0.132  & -0.091   \\ 
        22 & -0.183  & 0.191  & 0.194  & -0.283  & 0.208  & -0.218  & -0.011  & -0.137   \\ 
        23 & -0.182  & 0.192  & 0.195  & -0.299  & 0.211  & -0.222  & -0.015  & -0.135   \\ 
        24 & -0.167  & 0.191  & 0.195  & -0.275  & 0.192  & -0.204  & -0.003  & -0.128   \\ 
        25 & -0.078  & 0.237  & 0.255  & -0.186  & 0.260  & -0.174  & 0.148  & -0.095   \\ 
        26 & -0.163  & 0.189  & 0.195  & -0.277  & 0.181  & -0.204  & -0.003  & -0.129   \\ 
        27 & -0.140  & 0.198  & 0.199  & -0.284  & 0.168  & -0.192  & 0.077  & -0.111   \\ 
        28 & -0.145  & 0.206  & 0.210  & -0.274  & 0.168  & -0.192  & 0.071  & -0.109   \\ 
        29 & -0.170  & 0.182  & 0.184  & -0.303  & 0.175  & -0.211  & -0.009  & -0.132   \\ 
        30 & -0.161  & 0.189  & 0.194  & -0.320  & 0.188  & -0.214  & -0.015  & -0.124   \\ 
        31 & -0.065  & 0.200  & 0.222  & -0.158  & 0.227  & -0.143  & 0.122  & -0.083   \\ 
        32 & -0.162  & 0.181  & 0.182  & -0.315  & 0.142  & -0.216  & -0.005  & -0.135   \\ 
        33 & -0.164  & 0.175  & 0.177  & -0.314  & 0.141  & -0.215  & -0.005  & -0.135   \\ 
        34 & -0.144  & 0.194  & 0.204  & -0.274  & 0.197  & -0.197  & 0.001  & -0.117   \\ 
        35 & -0.101  & 0.180  & 0.193  & -0.234  & 0.183  & -0.182  & 0.088  & -0.111   \\ 
        36 & -0.054  & 0.189  & 0.270  & -0.170  & -0.001  & -0.147  & -0.033  & -0.079   \\ 
        37 & -0.095  & 0.216  & 0.231  & -0.224  & 0.250  & -0.192  & 0.141  & -0.108   \\ 
        38 & -0.056  & 0.197  & 0.213  & -0.245  & 0.194  & -0.155  & 0.102  & -0.089   \\ 
        39 & -0.086  & 0.204  & 0.215  & -0.168  & 0.214  & -0.155  & 0.107  & -0.089   \\ 
        40 & -0.110  & 0.185  & 0.199  & -0.233  & 0.211  & -0.187  & 0.108  & -0.116   \\ 
        41 & -0.082  & 0.201  & 0.214  & -0.192  & 0.192  & -0.161  & 0.097  & -0.092   \\ 
        42 & -0.091  & 0.204  & 0.221  & -0.193  & 0.230  & -0.168  & 0.120  & -0.100   \\ 
        43 & -0.012  & 0.051  & 0.057  & -0.029  & -0.089  & -0.036  & -0.126  & -0.123   \\ 
        44 & -0.069  & 0.226  & 0.251  & -0.181  & 0.192  & -0.157  & 0.082  & -0.082   \\ 
        45 & -0.146  & 0.166  & 0.170  & -0.247  & 0.203  & -0.212  & -0.011  & -0.140   \\ 
        \hline\hline
        \label{tab:peng_b3lyp}
\end{longtable}

\begin{longtable}{ccc p{1.2cm}p{1.7cm}p{1.2cm}p{1.2cm}p{1.2cm}p{1.2cm}}
\caption{Calculated $\Delta E_{\mathrm{ST}}$ (in eV) with PBE0/aug-cc-pVDZ for the INVEST-45 subset.} \\
\hline
\hline

        ~ & ~ & ~ & ~ & ~ & ALDA0 & ~ & MCOL&   \\ \cmidrule(r){6-7} \cmidrule(r){8-9}
        Molecule & CC2 & TDA & TD-DFT & USF-TD-DFT & USF-TDA & XSF-TDA & USF-TDA & XSF-TDA  \\ 
        \hline
\endfirsthead
\multicolumn{9}{c}{\tablename~\thetable{} (continued)} \\
\hline
\hline

        ~ & ~ & ~ & ~ & ~ & ALDA0 & ~ & MCOL&   \\ \cmidrule(r){6-7} \cmidrule(r){8-9}
        Molecule & CC2 & TDA & TD-DFT & USF-TD-DFT & USF-TDA & XSF-TDA & USF-TDA & XSF-TDA  \\ 
        \hline
\endhead
\hline
\multicolumn{9}{r}{Continued on next page} \\
\endfoot
\endlastfoot

        1 & -0.215  & 0.225  & 0.262  & -0.384  & 0.302  & -0.244  & -0.013  & -0.127   \\ 
        2 & -0.100  & 0.268  & 0.309  & -0.300  & 0.276  & -0.173  & 0.136  & -0.075   \\ 
        3 & -0.117  & 0.270  & 0.313  & -0.412  & 0.285  & -0.199  & 0.154  & -0.080   \\ 
        4 & -0.192  & 0.222  & 0.245  & -0.427  & 0.269  & -0.248  & -0.021  & -0.134   \\ 
        5 & -0.145  & 0.223  & 0.241  & -0.479  & 0.189  & -0.233  & 0.006  & -0.122   \\ 
        6 & -0.093  & 0.266  & 0.303  & -0.289  & 0.285  & -0.171  & 0.165  & -0.050   \\ 
        7 & -0.076  & 0.272  & 0.315  & -0.260  & 0.337  & -0.197  & 0.195  & -0.085   \\ 
        8 & -0.135  & 0.237  & 0.271  & -0.370  & 0.116  & -0.208  & -0.002  & -0.103   \\ 
        9 & -0.138  & 0.246  & 0.282  & -0.319  & 0.287  & -0.199  & 0.146  & -0.100   \\ 
        10 & -0.093  & 0.241  & 0.274  & -0.365  & 0.132  & -0.199  & 0.014  & -0.105   \\ 
        11 & -0.137  & 0.232  & 0.265  & -0.317  & 0.281  & -0.197  & 0.023  & -0.108   \\ 
        12 & -0.093  & 0.228  & 0.257  & -0.290  & 0.278  & -0.190  & 0.135  & -0.086   \\ 
        13 & -0.090  & 0.237  & 0.277  & -0.276  & 0.307  & -0.189  & 0.164  & -0.096   \\ 
        14 & -0.095  & 0.218  & 0.252  & -0.335  & 0.263  & -0.205  & 0.131  & -0.097   \\ 
        15 & -0.089  & 0.230  & 0.272  & -0.283  & 0.277  & -0.197  & 0.150  & -0.086   \\ 
        16 & -0.321  & 0.118  & 0.129  & -0.708  & 0.045  & -0.379  & -0.108  & -0.250   \\ 
        17 & -0.156  & 0.221  & 0.251  & -0.418  & 0.265  & -0.223  & 0.005  & -0.119   \\ 
        18 & -0.133  & 0.240  & 0.274  & -0.387  & 0.270  & -0.208  & 0.003  & -0.105   \\ 
        19 & -0.152  & 0.215  & 0.241  & -0.411  & 0.123  & -0.229  & -0.005  & -0.129   \\ 
        20 & -0.069  & 0.222  & 0.266  & -0.269  & 0.281  & -0.168  & 0.137  & -0.080   \\ 
        21 & -0.080  & 0.233  & 0.287  & -0.224  & 0.320  & -0.185  & 0.031  & -0.090   \\ 
        22 & -0.183  & 0.212  & 0.233  & -0.433  & 0.267  & -0.249  & -0.020  & -0.139   \\ 
        23 & -0.182  & 0.214  & 0.237  & -0.478  & 0.267  & -0.255  & -0.026  & -0.138   \\ 
        24 & -0.167  & 0.212  & 0.235  & -0.425  & 0.255  & -0.232  & -0.009  & -0.132   \\ 
        25 & -0.078  & 0.261  & 0.302  & -0.305  & 0.325  & -0.212  & 0.188  & -0.092   \\ 
        26 & -0.163  & 0.210  & 0.234  & -0.432  & 0.233  & -0.233  & -0.014  & -0.134   \\ 
        27 & -0.140  & 0.224  & 0.243  & -0.507  & 0.221  & -0.223  & 0.002  & -0.117   \\ 
        28 & -0.145  & 0.230  & 0.253  & -0.467  & 0.211  & -0.221  & -0.005  & -0.112   \\ 
        29 & -0.170  & 0.203  & 0.221  & -0.467  & 0.102  & -0.243  & -0.025  & -0.139   \\ 
        30 & -0.161  & 0.213  & 0.240  & -0.577  & 0.243  & -0.249  & -0.034  & -0.129   \\ 
        31 & -0.065  & 0.221  & 0.269  & -0.232  & 0.119  & -0.169  & 0.024  & -0.080   \\ 
        32 & -0.162  & 0.202  & 0.217  & -0.519  & 0.106  & -0.251  & -0.022  & -0.145   \\ 
        33 & -0.164  & 0.196  & 0.212  & -0.513  & 0.105  & -0.248  & -0.023  & -0.145   \\ 
        34 & -0.144  & 0.217  & 0.250  & -0.449  & 0.111  & -0.226  & 0.001  & -0.123   \\ 
        35 & -0.101  & 0.203  & 0.234  & -0.371  & 0.245  & -0.218  & -0.002  & -0.110   \\ 
        36 & -0.054  & 0.215  & 0.422  & -0.199  & 0.026  & -0.168  & -0.050  & -0.071   \\ 
        37 & -0.095  & 0.239  & 0.275  & -0.373  & 0.311  & -0.235  & 0.177  & -0.109   \\ 
        38 & -0.056  & 0.221  & 0.259  & -0.393  & 0.067  & -0.188  & -0.016  & -0.092   \\ 
        39 & -0.086  & 0.226  & 0.255  & -0.264  & 0.274  & -0.186  & 0.137  & -0.085   \\ 
        40 & -0.110  & 0.206  & 0.237  & -0.357  & 0.088  & -0.224  & -0.009  & -0.117   \\ 
        41 & -0.082  & 0.224  & 0.262  & -0.306  & 0.255  & -0.192  & 0.134  & -0.087   \\ 
        42 & -0.091  & 0.226  & 0.264  & -0.314  & 0.294  & -0.202  & 0.154  & -0.101   \\ 
        43 & -0.012  & 0.062  & 0.071  & -0.042  & 0.062  & 0.050  & 0.006  & -0.125   \\ 
        44 & -0.069  & 0.256  & 0.307  & -0.267  & 0.244  & -0.168  & 0.104  & -0.064   \\ 
        45 & -0.146  & 0.188  & 0.205  & -0.352  & 0.266  & -0.237  & -0.023  & -0.135   \\ 
        \hline\hline
    \label{tab:peng_pbe0}
\end{longtable}

\begin{longtable}{ccccc}
\caption{CV(1) percentage (\%) for the S$_1$ state in XSF-TDA calculations for the INVEST-45 subset.}\\
\hline
\hline
        ~ & B3LYP & ~ & PBE0 & ~ \\ \cmidrule(r){2-3} \cmidrule(r){4-5}
        Molecule & ALDA0 & MCOL& ALDA0 & MCOL\\ 
        \hline
\endfirsthead
\multicolumn{5}{c}{\tablename~\thetable{} (continued)} \\
\hline
\hline
        ~ & B3LYP & ~ & PBE0 & ~ \\ \cmidrule(r){2-3} \cmidrule(r){4-5}
        Molecule & ALDA0 & MCOL& ALDA0 & MCOL\\ 
        \hline
\endhead
\hline
\multicolumn{5}{r}{Continued on next page} \\
\endfoot
\endlastfoot

        1 & 4.576  & 3.289  & 5.224  & 3.601  \\ 
        2 & 4.490  & 3.228  & 5.198  & 3.601  \\ 
        3 & 5.550  & 4.085  & 6.391  & 4.577  \\ 
        4 & 4.606  & 3.288  & 5.305  & 3.622  \\ 
        5 & 5.270  & 3.815  & 6.101  & 4.270  \\ 
        6 & 5.659  & 4.213  & 6.350  & 4.633  \\ 
        7 & 4.301  & 3.049  & 4.941  & 3.330  \\ 
        8 & 5.492  & 4.039  & 6.299  & 4.519  \\ 
        9 & 4.720  & 3.390  & 5.425  & 3.755  \\ 
        
        10 & 5.381  & 3.934  & 6.191  & 4.407  \\ 
        11 & 4.846  & 3.488  & 5.575  & 3.877  \\ 
        12 & 4.158  & 2.952  & 4.781  & 3.226  \\ 
        13 & 4.409  & 3.131  & 5.071  & 3.431  \\ 
        14 & 4.492  & 3.205  & 5.151  & 3.505  \\ 
        15 & 4.531  & 3.257  & 5.209  & 3.578  \\ 
        16 & 5.200  & 3.698  & 5.977  & 4.043  \\ 
        17 & 5.172  & 3.748  & 5.951  & 4.164  \\ 
        18 & 5.272  & 3.843  & 6.073  & 4.289  \\ 
        19 & 4.895  & 3.510  & 5.655  & 3.904  \\ 
        20 & 4.506  & 3.202  & 5.202  & 3.530  \\ 
        21 & 4.528  & 3.251  & 5.204  & 3.582  \\ 
        22 & 4.898  & 3.513  & 5.643  & 3.887  \\ 
        23 & 5.237  & 3.789  & 6.022  & 4.197  \\ 
        24 & 4.997  & 3.597  & 5.767  & 4.000  \\ 
        25 & 4.656  & 3.324  & 5.347  & 3.643  \\ 
        26 & 4.693  & 3.347  & 5.436  & 3.715  \\ 
        27 & 5.676  & 4.159  & 6.550  & 4.665  \\ 
        28 & 5.130  & 3.719  & 5.933  & 4.155  \\ 
        29 & 5.298  & 3.845  & 6.115  & 4.294  \\ 
        30 & 5.840  & 4.292  & 6.723  & 4.797  \\ 
        31 & 4.721  & 3.413  & 5.425  & 3.777  \\ 
        32 & 5.272  & 3.789  & 6.105  & 4.230  \\ 
        33 & 5.193  & 3.735  & 6.018  & 4.170  \\ 
        34 & 5.620  & 4.108  & 6.479  & 4.587  \\ 
        35 & 4.821  & 3.452  & 5.532  & 3.789  \\ 
        36 & 5.008  & 3.703  & 5.571  & 3.975  \\ 
        37 & 5.049  & 3.620  & 5.798  & 3.975  \\ 
        38 & 5.339  & 3.931  & 6.129  & 4.357  \\ 
        39 & 4.228  & 3.013  & 4.890  & 3.315  \\ 
        40 & 4.915  & 3.545  & 5.639  & 3.899  \\ 
        41 & 4.668  & 3.332  & 5.369  & 3.658  \\ 
        42 & 4.694  & 3.348  & 5.406  & 3.683  \\ 
        43 & 2.365  & 1.455  & 2.931  & 1.736  \\ 
        44 & 3.854  & 2.737  & 4.363  & 2.957  \\ 
        45 & 4.163  & 2.932  & 4.774  & 3.180  \\ 
        \hline\hline
    \label{tab:peng_cv1}
\end{longtable}

\begin{longtable}{cccc}
\caption{Calculated $\Delta E_{\mathrm{ST}}$ (in eV) with SC-NEVPT2(8,8)/cc-pVDZ, ADC(2)/aug-cc-pVDZ, and CC2/aug-cc-pVDZ for the INVEST-45 subset.} \\
\hline
\hline
        Molecule & SC-NEVPT2(8,8)\cite{peng2025} & ADC(2) & CC2  \\
        \hline
\endfirsthead
\multicolumn{4}{c}{\tablename~\thetable{} (continued)} \\
\hline
\hline
        Molecule & SC-NEVPT2(8,8)\cite{peng2025} & ADC(2) & CC2  \\
        \hline
\endhead
\hline
\multicolumn{4}{r}{Continued on next page} \\
\endfoot
\endlastfoot
        1 & -0.224  & -0.222  & -0.215   \\ 
        2 & -0.019  & -0.109  & -0.100   \\ 
        3 & -0.133  & -0.135  & -0.117   \\ 
        4 & -0.235  & -0.199  & -0.192   \\ 
        5 & -0.105  & -0.147  & -0.145   \\ 
        6 & -0.079  & -0.109  & -0.093   \\ 
        7 & -0.078  & -0.086  & -0.076   \\ 
        8 & -0.025  & -0.140  & -0.135   \\ 
        9 & -0.126  & -0.144  & -0.138   \\ 
        10 & -0.062  & -0.096  & -0.093   \\ 
        11 & -0.199  & -0.141  & -0.137   \\ 
        12 & -0.117  & -0.096  & -0.093   \\ 
        13 & -0.127  & -0.096  & -0.090   \\ 
        14 & -0.092  & -0.102  & -0.095   \\ 
        15 & -0.026  & -0.097  & -0.089   \\ 
        16 & -0.025  & -0.316  & -0.321   \\ 
        17 & -0.130  & -0.166  & -0.156   \\ 
        18 & -0.021  & -0.141  & -0.133   \\ 
        19 & -0.119  & -0.158  & -0.152   \\ 
        20 & -0.110  & -0.066  & -0.069   \\ 
        21 & -0.067  & -0.087  & -0.080   \\ 
        22 & -0.054  & -0.188  & -0.183   \\ 
        23 & -0.183  & -0.190  & -0.182   \\ 
        24 & -0.189  & -0.173  & -0.167   \\ 
        25 & -0.190  & -0.087  & -0.078   \\ 
        26 & -0.142  & -0.170  & -0.163   \\ 
        27 & -0.105  & -0.143  & -0.140   \\ 
        28 & -0.159  & -0.153  & -0.145   \\ 
        29 & -0.151  & -0.174  & -0.170   \\ 
        30 & -0.141  & -0.171  & -0.161   \\ 
        31 & -0.035  & -0.070  & -0.065   \\ 
        32 & -0.289  & -0.165  & -0.162   \\ 
        33 & -0.195  & -0.167  & -0.164   \\ 
        34 & -0.129  & -0.156  & -0.144   \\ 
        35 & -0.106  & -0.110  & -0.101   \\ 
        36 & -0.098  & -0.076  & -0.054   \\ 
        37 & -0.172  & -0.102  & -0.095   \\ 
        38 & -0.067  & -0.071  & -0.056   \\ 
        39 & -0.070  & -0.087  & -0.086   \\ 
        40 & -0.111  & -0.119  & -0.110   \\ 
        41 & -0.049  & -0.085  & -0.082   \\ 
        42 & -0.108  & -0.096  & -0.091   \\ 
        43 & -0.108  & -0.010  & -0.012   \\ 
        44 & -0.171  & -0.090  & -0.069   \\ 
        45 & -0.149  & -0.152  & -0.146   \\
        \hline\hline
    \label{tab:peng_ref}
\end{longtable}
}

For reproducibility, we present an example for using the CC2 method in Q‑Chem as follows:
\begin{verbatim}
$molecule
 0 1
 C  2.104057  1.274661  5e-05
 N  1.4406  2.448732  -0.000252
 C  0.088078  2.457887  5.1e-05
 C  -0.735933  1.307833  0.000763
 C  -2.152039  1.243466  -0.000195
 C  -2.820852  0.001229  -0.000448
 C  -2.153142  -1.241621  0.000156
 C  -0.737165  -1.307147  0.00064
 B  -0.004808  2e-06  0.000756
 C  1.487618  -0.000679  2.1e-05
 C  2.102941  -1.276497  -0.000289
 N  1.438422  -2.449987  -0.000238
 C  0.08591  -2.457932  -8.4e-05
 H  3.19967  1.379957  0.000399
 H  -0.348542  3.468416  -0.001354
 H  -2.764406  2.15474  -0.000904
 H  -3.913207  0.001707  -0.001195
 H  -2.7663  -2.15235  0.000362
 H  3.198447  -1.382827  -0.000728
 H  -0.351617  -3.468079  -0.000918
$end

$rem
 JOBTYPE SP
 METHOD cc2
 AUX_BASIS rimp2-vdz
 CCMAN2 true
 BASIS aug-cc-pvdz
 EE_SINGLETs [3]
 EE_TRIPLETS [3]
 MEM_TOTAL 128000
 CC_MEMORY 100000
 SYMMETRY false
 SYM_IGNORE true
$end
\end{verbatim}

\clearpage
\bibliography{reference}